\documentclass[journal]{IEEEtran}

\usepackage{xcolor}
\usepackage{psfrag}
\usepackage{pgfplots}
\usepackage{resizegather}
\usepackage{graphics} 
\usepackage{epsfig} 
\usepackage{times} 
\usepackage{amsmath} 
\usepackage{amssymb}  
\usepackage{mathtools}
\usepackage{array}
\usepackage{bm}
\usepackage{tikz}
\usetikzlibrary{calc,arrows}
\usepackage{pgfplots}
\usepackage{subcaption}
\usepackage{etex}
\usepackage{siunitx}
\usepackage{url}
\usepackage{arydshln} 
\usepackage[ruled,vlined,linesnumbered]{algorithm2e}
 
\usepackage{enumitem} 

\usepackage[bottom]{footmisc}

\usepackage[noend]{algpseudocode}
\usepackage{stackengine}
\usepackage{relsize}

\usepackage{multirow}

\makeatletter
\def\BState{\State\hskip-\ALG@thistlm}
\makeatother

\DeclareMathOperator*{\diagonal}{diag}

\def\qed {{
\parfillskip=0pt 
\widowpenalty=10000 
\displaywidowpenalty=10000 
\finalhyphendemerits=0 
\leavevmode 
\unskip 
\nobreak 
\hfil 
\penalty50 
\hskip.2em 
\null 
\hfill 
$\blacksquare$
\par}}

\newtheorem{myrem}{Remark}
\newtheorem{myexp}{Example}
\newtheorem{mythe}{Theorem}

\newtheorem{mydef}{Definition}
\newtheorem{myprop}{Proposition}
\newtheorem{mycor}{Corollary}

\newcommand{\iter}[0]{\kappa}

\newcommand{\bma}[2]       {\left[\begin{array}{#1}#2\end{array}\right]}	
\newcommand{\mrm}[1]       {\mathrm{#1}}

\newcommand{\kronecker}[2] {{#1}\otimes{#2}}
\newcommand{\bd}[1]   {{#1}^d}
\newcommand{\obd}[1]   {{#1}^i}
\newcommand\norm[1]{\left\lVert#1\right\rVert}

\newcommand{\gammainv}[0]{\nu} 

\newcommand{\aalpha}[0]   {\alpha}
\newcommand{\intercN}[0]  {\beta}
\newcommand{\gl}[0]{l}

\newcommand{\perfo}[0]{{z}} 
\newcommand{\perfi}[0]{{w}} 
\newcommand{\intercPo}[0]{q} 
\newcommand{\intercPi}[0]{p} 
\newcommand{\intercKo}[0]{q^K} 
\newcommand{\intercKi}[0]{p^K} 
\newcommand{\interco}[0]{q^c} 
\newcommand{\interci}[0]{p^c} 

\newcommand{\y}[0]{s}

\newcommand\barbelow[1]{\stackunder[1.2pt]{$#1$}{\rule{.8ex}{.075ex}}}

\newcommand{\pattern}[0]{P}
\newcommand{\permute}[0]{\mathcal{P}}

\newcommand{\Plant}[0]{G}
\newcommand{\Plantclp}[0]{\mathcal{G}}
\newcommand{\plant}[1]{{#1}^{G}}
\newcommand{\clp}[1]{{#1}^{c}}
\newcommand{\cont}[1]{{#1}^{K}}
\newcommand{\contmirror}[1]{{#1}^{KM}}
\newcommand{\plantmirror}[1]{{#1}^{GM}}

\newcommand{\graph}[0]{\mathcal{T}}

\newcommand{\xK}[0]{x^K}
\newcommand{\xP}[0]{x}
\newcommand{\xc}[0]{x^{c}}
\newcommand{\dotxK}[0]{\dot{x}^K}
\newcommand{\dotxP}[0]{\dot{x}}
\newcommand{\dotxc}[0]{\dot{x}^{c}}
\newcommand{\AK}[0]{A^K}
\newcommand{\BK}[0]{B^K}
\newcommand{\CK}[0]{C^K}
\newcommand{\DK}[0]{D^K}

\newcommand{\smallQ}[1]{\tilde{#1}}

\newcommand{\diagIndtwo}[3]      {\mrm{\diagonal}_{#1}^{#2}\!\left(#3\right)}		
\newcommand{\concatIndtwo}[3]      {\mrm{concat}_{#1}^{#2}\!\left(#3\right)}		
\newcommand{\trans}        {\top}																				  
		
\newcommand{\controrig} {K}

\newcommand{\LFTu} {\mathcal{F}_{\text{u}}}	
\newcommand{\LFTl} {\mathcal{F}_{\text{l}}}

\definecolor{ForestGreen}{rgb}{0.0, 0.27, 0.13}
\definecolor{LimeGreen}{rgb}{0.2, 0.8, 0.2}
\definecolor{Blue}{rgb}{0.0, 0.0, 1.0}
\definecolor{ProcessBlue}{rgb}{0.0, 0.72, 0.92}
\definecolor{Melon}{rgb}{0.99, 0.74, 0.71}
\definecolor{Peach}{rgb}{1.0, 0.9, 0.71}
\definecolor{darktangerine}{rgb}{1.0, 0.66, 0.07}
\definecolor{Red}{rgb}{0.8, 0.0, 0.0}

\title{\LARGE \bf
Distributed Control Design for Heterogeneous Interconnected Systems
}

\author{Yvonne R.\ St\"urz, Annika Eichler and Roy S.\ Smith%
\thanks{Yvonne R.\ St\"urz is with the Model Predictive Control Laboratory, UC Berkeley, 2505 Hearst Avenue, CA 94709, USA. 
	Annika Eichler is with the DESY, Hamburg, Germany. 
	 Roy S.\ Smith is with the Automatic Control Laboratory, ETH Zurich, Physikstrasse 3, 8092 Zurich, Swizerland. Email addresses: %
        \tt\small y.stuerz@berkeley.edu, 
        annika.eichler@desy.de, 
        rsmith@control.ee.ethz.ch
    }}%

\begin{document}

\maketitle
\thispagestyle{empty}
\pagestyle{empty}

\begin{abstract}
	This paper presents scalable controller synthesis methods for heterogeneous and partially heterogeneous systems. 
	First, heterogeneous systems composed of different subsystems that are interconnected over a directed graph are considered. 
	Techniques from robust and gain-scheduled controller synthesis are employed, in particular the full-block S-procedure, to deal with the decentralized system part in a nominal condition and with the interconnection part in a multiplier condition. 
	Under some structural assumptions, we can decompose the synthesis conditions into conditions that are the size of the individual subsystems. 
	To solve these decomposed synthesis conditions that are coupled only over neighboring subsystems, we propose a distributed method based on the alternating direction method of multipliers. It only requires nearest-neighbor communication and no central coordination is needed. 
	Then, a new classification of systems is introduced that consists of groups of homogeneous subsystems with different interconnection types. This classification includes heterogeneous systems as the most general and homogeneous systems as the most specific case. Based on this classification, we show how the interconnected system model and the decomposed synthesis conditions can be formulated in a more compact way. 
	The computational scalability of the presented methods with respect to a growing number of subsystems and interconnections is analyzed, and the results are demonstrated in numerical examples. 	
\end{abstract}

\section{Introduction}

Distributed control of interconnected systems has been an active field of research \cite{DAndrea2003, Langbort2004}. Applications range from multi-agent systems \cite{Murray2007}, such as formation control of spacecraft \cite{Smith2007}, cooperative robotic manipulation \cite{Stuerz2017a}, to automated highways \cite{Alam2011}. 
Distributed systems can be modeled as interconnected systems with a decentralized part and an interconnection part. 
This has been introduced as a general representation based on a linear fractional representation (LFR) of the system by \cite{Gahinet1995} and \cite{Scorletti1997} for gain-scheduled and decentralized controller design, respectively, and subsequently used in \cite{Langbort2004} for distributed control design. 
In \cite{DAndrea2003, Dullerud2004, Langbort2003} similar models are used to represent spatially interconnected systems. 
In these works, distributed controllers of heterogeneous systems over undirected graphs are synthesized, where the interconnection structure of the controller is assumed to be the same as that of the plant. 
In \cite{Rice2009}, spatially-varying interconnected systems distributed in a spatial dimension are considered, for which distributed controllers are designed. 	
The work \cite{Borrelli2008} considers distributed controller design for identical dynamically decoupled subsystems, which are only coupled through the performance objective. 

For the analysis and distributed controller synthesis for interconnected systems, tools from robust and gain-scheduled controller synthesis \cite{Rantzer1997, Meinsma2000} have been successfully applied. 
The full block S-procedure \cite{Scherer2001} has been employed in \cite{Massioni2009, Massioni2014} to the special case of systems of homogeneous subsystems and groups of homogeneous subsystems interconnected over undirected graphs. Under structural assumptions on the Lyapunov and multiplier matrices, a signal transformation was introduced in \cite{Massioni2010} to decompose the synthesis equations into smaller ones for the individual subsystems. 
In \cite{Hoffmann2015} a congruence transformation is proposed instead of the signal transformation to cope with time-varying and heterogeneous subsystems. 
In \cite{Hoffmann2015}, groups of homogeneous subsystems are considered, which need to be interconnected by undirected interactions within the groups and by directed ones between the groups. \cite{Hoffmann2016} introduces a transformation which can deal with directed graphs and groups of heterogeneous subsystems. The approaches in \cite{Massioni2014, Hoffmann2015, Hoffmann2016} are however restricted to equal interconnections within the groups of homogeneous subsystems. 
In \cite{Stuerz2018a}, an approach to deal with the more general case of groups of homogeneous subsystems with different interconnections is proposed.

In \cite{Langbort2004}, in addition to structural assumptions on the Lyapunov and multiplier conditions, DG scalings are imposed, which in general introduces further conservatism. Together with the structure of the undirected interconnection graph of the plant and the controller, this leads to synthesis equations which are pairwise coupled per interconnection, and which need to be solved in a centralized way. 
In \cite{Langbort2004a}, a distributed algorithm for the controller synthesis is proposed which is based on a primal decomposition and subgradient methods. Coupling in the synthesis equations from \cite{Langbort2004} occurs between each two subsystems connected over an edge, and both of these subsystems compute subgradient information for the distributed synthesis algorithm, where the updates occur in a sequential way. 
The work in \cite{Viccione2009} makes use of a similar framework as in \cite{Langbort2004} for synthesizing a controller with an interconnection structure which does not need to replicate the one of the plant. The full block S-procedure is employed. However, no distributed synthesis methods are proposed.

In \cite{Zheng2018a} a different approach for distributed controller synthesis is presented, where instead of the full-block S-procedure, a chordal decomposition is used to decompose the centralized synthesis equations, and a sequential design approach is taken. However, if a performance criterion is to be optimized, this approach can in general lead to conservatism or infeasibility. 
   
In the area of distributed control and optimization, the Alternating Direction Method of Multipliers (ADMM) has become a widely used method \cite{Boyd2010}. It combines the advantages of dual decomposition, such as parallel and distributed computation, with the advantages of the method of multipliers, such as convexification. 
ADMM has been used in numerous applications, for example in electrical power systems 
\cite{Molzahn2017, Braun2018, Guo2018}, 
in distributed system and parameter identification for large-scale systems \cite{Hansson2014, Stuerz2016b}, 
in optimal traffic flow problems \cite{Ba2015}, 
\cite{Dhingra2014} in sensor and actuator selection, and in distributed reinforcement learning \cite{Graf2019}. 
In \cite{Alonso2019}, ADMM is used in a distributed model predictive control scheme based on the system level synthesis.

The contributions of this paper are the following: 

We propose a scalable distributed controller synthesis for large-scale heterogeneous systems, where the subsystems can be interconnected over a directed graph. 
The synthesized distributed controller may have a directed interconnection topology that can be different from the topology of the plant. 

The scalability of the proposed controller synthesis method is achieved in two aspects. 
First, through applying the full block S-procedure we decompose the controller synthesis equations. 
This is achieved based on structural assumptions on the Lyapunov and multiplier equations, and through congruence transformations. 
Allowing for structured full block multipliers potentially reduces conservatism compared to imposing D or DG scalings. 
The resulting decomposed conditions are of the size of the individual subsystems and are pairwise coupled over the edges of the interconnection topology. 

To solve the decomposed synthesis equations, we propose a distributed method which is fully parallelizable and which involves only bidirectional nearest-neighbor communication and no central coordinator. 
The distributed synthesis is based on ADMM. 
By choosing the variable splitting in a specific way, the ADMM algorithm can be simplified to two steps with only nearest neighbor communication. 
This simplification has been introduced in \cite{Mateos2010} for the distributed Lasso problem and has been extended towards more complex cost functions in \cite{Chang2015}. In \cite{Banjac2018}, the algorithm was generalized to deal with conic constraints in the consensus couplings.

We further introduce a new system classification, referred to as $\aalpha$-$\intercN$-heterogeneous systems, that consist of $\aalpha$ groups of homogeneous subsystems with $\intercN$ different interconnection types. 
This is an extension to the model considered in \cite{Massioni2014}, where all interconnections are required to be of the same type.  
$\aalpha$-$\intercN$-heterogeneous systems include heterogeneous systems as the most general, and homogeneous systems as the most specific case. 
We show how, based on this classification, the interconnected system model can be transformed to a more compact form. 
For small values of $\aalpha$ and $\intercN$ less conservatism is introduced by a controller synthesis based on the more compact model that captures the information of the interconnection topology of the subsystems.

The paper is structured as follows. 
After a paragraph on the notation, we introduce the model of interconnected systems in Section~\ref{sec:model}, the controller model and the closed loop system in Section~\ref{sec:controlclosedloop}, and give a transformation of a general system to an interconnected system model in Section~\ref{sec:PerformanceAugmentation}. Then, we present the decomposition of the controller synthesis equations for heterogeneous systems and a distributed design method based on the alternating direction method of multipliers (ADMM) in Sections~\ref{sec:DecomposedEquations} and \ref{sec:DistributedDesign}, before presenting special classes of interconnected system models and their simplified decomposed controller synthesis in Section~\ref{sec:SpecialCases}. The last two sections illustrate the results by numerical examples and conclude the paper.

\subsection{Notation}
We denote a block-diagonal matrix $D$ of submatrices $D_1,...,D_N$ by $D = \diagIndtwo{i=1}{N}{D_i}$, and a matrix $C$ composed of submatrices $C_1,...,C_N$ as block-rows is denoted by $\concatIndtwo{i=1}{N}{C_i}$. The $n \times n$-identity matrix is denoted by $I_n$ and the $n \times m$ matrix of all zeros as $0_{n \times m}$. If clear from the context, the indices are dropped. $\hat{I}_{\{ i \}}$ is defined as a square matrix of appropriate dimensions of all zeros except that the $(i,i)$-entry is one, and $e_{j}$ is the $j$-th unit vector. The spectrum of the matrix $M$, i.e., the set of its eigenvalues, is denoted by $\mathrm{spec}(M)$. Minimum and maximum eigenvalues are denoted by $\lambda_{\mathrm{min}}(\cdot)$ and  $\lambda_{\mathrm{max}}(\cdot)$. Similarly, minimum and maximum singular values are denoted by $\sigma_{\mathrm{min}}(\cdot)$ and  $\sigma_{\mathrm{max}}(\cdot)$. 
The Kronecker product is denoted by $\kronecker{}{}$ and the Moore-Penrose pseudo-inverse by $\dag$. 
Given a complex valued matrix $M=\small \bma{@{\,}c@{\,\;}c@{\,}}{M_1 & M_2\\ M_3 & M_4}$ and $P$ of appropriate dimensions, then the lower and upper Linear Fractional Transformation (LFT) are defined as $\mathcal{F}_\text{l}(M,P) =M_1+M_2P(I-M_4)^{-1}M_3$ and $\mathcal{F}_\text{u}(M,P) =M_4+M_3P(I-M_1)^{-1}M_2$, respectively. We use the symbol $\star$ to simplify expressions as $M_1^\trans M_2 M_1$, i.e., $\star^\trans M_2 M_1 = M_1^\trans M_2 M_1$. 
$\text{vect}(\cdot)$ is an operator which forms a vector out of the matrix argument. 
System matrices $M$ belonging to the specific channel $w$ to $z$ of subsystem $i$ are denoted by $M_{zw,i}$. The superscripts $G$ and $K$ are used to indicate the graph topology of the system $G$ or the controller $K$. For signals, no superscript, or the superscripts $K$ and $c$ are used to indicate that the signal belongs to the plant, controller, or closed-loop, respectively. The dimension of a signal $x$ is denoted by $n_x$. 

\section{Interconnected Systems}
\label{sec:model}

We consider a system of $N$ different LTI subsystems which are interconnected by an arbitrary directed graph. This section presents the interconnected system model.

\subsection{Graph Structure}
\label{subsec:Graphstructure}

We consider a connected graph $\plant{\graph}=\{\mathcal{N},\plant{\mathcal{E}}\}$ where its vertices are 
$N$ possibly different finite dimensional LTI subsystems, associated with the node set $\mathcal{N}=\{1,…,N\}$, which is the index set of all subsystems $\Plant_i, i \in \mathcal{N}$. 
We further define the set of directed edges $\plant{\mathcal{E}} := \{(i,k)\}$ for all pairs $(i,k)$ where subsystem $i$ is influenced by subsystem $k$. 
The interconnection topology of the subsystems is captured by the interconnection matrix $\plant{\pattern}$, 
which is an $N \times N$ matrix of all zeros except for non-zero entries in the places corresponding to interconnections between subsystems $i$ and $k$, i.e., ${\plant{\pattern}}_{ik}$ is non-zero.  

While the interconnection graph $\plant{\graph}$ capturing the interconnections between the subsystems may be directed, 
we will in Section~\ref{sec:DistributedDesign} assume that subsystems interconnected by an edge in $\plant{\mathcal{E}}$ can communicate in a bidirectional way during controller synthesis. 
To this end, and for the interconnection representation which will be intorduced in Section~\ref{subsec:interconn}, we introduce the mirror graph $\plantmirror{\graph}:=\{\mathcal{N},\plantmirror{\mathcal{E}}\}$ as the graph which completes $\plant{\graph}$ to an undirected one, i.e., for all directed edges $(i,k) \in \plant{\mathcal{E}}$ for which there does not exist an edge $(k,i) \in \plant{\mathcal{E}}$, there exists an edge $(k,i)$ in $\plantmirror{\mathcal{E}}$. The interconnection matrix $\plantmirror{\pattern}$ is defined for $\plantmirror{\graph}$ analogously to $\plant{\pattern}$ for $\plant{\graph}$.

\subsection{Interconnected State Space Representations}
The subsystems $G_i$ admit the following state space representations. 
\begin{equation}
\label{eq:interconn_openloop_i}
\begin{aligned}
\Plant_i & : 
\begin{cases}
\bma{@{}c@{}}{ 
	\dotxP_i \\
	\hdashline 
	y_i \\
	\hdashline 
	\perfo_i \\
	\hdashline
	\intercPo_i \\
} 
= 
\bma{@{}c@{\,\,\,\,\,\,} : c@{\,\,\,\,\,\,}  : c@{\,\,\,\,\,\,} : c@{}}{ 
	{A}_i    & {B}_{ u,i} & {B}_{\perfi,i} & {B}_{\intercPi,i} \\
	\hdashline
	{C}_{y,i }  & 0 & {D}_{y \perfi,i} & {D}_{y\intercPi,i}  \\
	\hdashline
	{C}_{\perfo,i}  & {D}_{\perfo u,i} & {D}_{\perfo \perfi,i} & {D}_{\perfo \intercPi,i} \\
	\hdashline
	{C}_{\intercPo,i}   & 0 & {D}_{\intercPo \perfi,i} & 0 \\
}
\bma{@{}c@{}}{ 
	\xP_i \\
	\hdashline 
	u_i \\
	\hdashline 
	\perfi_i \\
	\hdashline
	\intercPi_i 
} , 
\end{cases} \\
\end{aligned}
\end{equation}
with the state vector $\xP_i \in \mathbb{R}^{n_{\xP i}}$, 
the control input and measured output, ${u}_i \in \mathbb{R}^{n_{ui}}$ and ${y}_i \in \mathbb{R}^{n_{yi}}$, the exogenous input and performance output, $\perfi_i \in \mathbb{R}^{n_{\perfi i}}$ and $\perfo_i \in \mathbb{R}^{n_{\perfo i}}$, and the interconnection signals $\intercPi_i \in \mathbb{R}^{n_{\intercPi i}}$ and $\intercPo_i \in \mathbb{R}^{n_{\intercPo i}}$, respectively. A transformation of a general distributed LTI system to the representation in  \eqref{eq:interconn_openloop_i} will be given in Section~\ref{sec:PerformanceAugmentation}, and a possible realization of the system matrices in \eqref{eq:interconn_openloop_i} is given in Appendix \ref{sub:LFTMatrices_hetero} in \eqref{eq:interconn_matrices_hetero}.

\subsection{Interconnection Relations}
\label{subsec:interconn}
\begin{figure}
	\input{figures/interconnections_convention.tex} 
	\includegraphics[width=0.99\columnwidth]{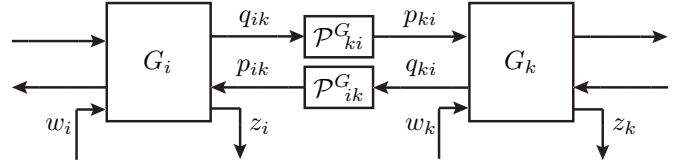} 
	\caption{Interconnection and performance inputs and outputs.  \label{fig:interconnections}}
\end{figure}

We define the set of neighboring subsystems of subsystem $i$, denoted by $\plant{\mathcal{N}}_{i}$, as the set of subsystems for which there exists an interconnection with subsystem $i$, i.e., for which there exists an edge $(i,k)$ in the union of edge sets $\plant{\mathcal{E}} \cup \plantmirror{\mathcal{E}}$. 
The interconnection signals $\intercPi_i$ and $\intercPo_i$ of the subsystems are further partitioned into 
\begin{equation} \label{eq:piqi}
\begin{aligned}
\intercPi_i &= \concatIndtwo{k \in \plant{\mathcal{N}}_i}{}{\intercPi_{ik}}, \\
\intercPo_i &= \concatIndtwo{k \in \plant{\mathcal{N}}_i}{}{\intercPo_{ik}},
\end{aligned}
\end{equation}
where $\intercPi_{ik} \in \mathbb{R}^{n_{\intercPi_{ik}}}$, $\intercPo_{ik} \in \mathbb{R}^{n_{\intercPo_{ik}}}$ are the incoming and outgoing interconnection signals of subsystem $i$ from and to subsystem $k$. 
The interconnection signals $\intercPi \in \mathbb{R}^{n_{\intercPi}}$ and $\intercPo \in \mathbb{R}^{n_{\intercPo}}$ of the system are then defined as 
\begin{equation} \label{eq:interc_signals_def}
\begin{aligned}
\intercPi &= \concatIndtwo{i \in \mathcal{N}}{}{\intercPi_{i}}, \\
\intercPo &= \concatIndtwo{i \in \mathcal{N}}{}{\intercPo_{i}}.
\end{aligned}
\end{equation}
Furthermore, we define the interconnection matrix $\plant{\permute}$ through the relation 
\begin{equation} \label{eq:P_hetero}
\intercPi = \plant{\permute} \intercPo. 
\end{equation} 
The entries in $\plant{\permute}$ are the elements $\plant{\permute}_{ik}$ of the individual interconnection relations
${\intercPi_{ik} =  \plant{\permute}_{ik} \, \intercPo_{ki}}$, which are shown in Figure~\ref{fig:interconnections}. 
In the case of ideal interconnections, these entries $\plant{\permute}_{ik}$ are identities of appropriate dimensions. 
An illustration of the definition of the interconnection signals and the interconnection matrix is given in Example~\ref{exp:trafo} in Section~\ref{sec:NumericalExample}.  

We denote the decentralized part of the plant, i.e., the ensemble of all $\Plant_i, ~\forall i=1,...,N$ as $\bd{\Plant}$. The interconnected system, denoted by $\Plant$, is then given by the decentralized part $\bd{\Plant}$ and the interconnection channel as follows 
\begin{equation}
\label{eq:interconn_openloop}
\begin{aligned}
{\Plant}: 
\begin{cases}
\bd{\Plant} = 
\diagIndtwo{i \in \mathcal{N}}{}{{\Plant}_i}, \\
\intercPi = \plant{\permute} \intercPo\,. 
\end{cases}
\end{aligned}
\end{equation}
Note that with the definition of $\intercPi_i$ and $\intercPo_i$ in \eqref{eq:piqi} over the interconnections $(i,k) \in \plant{\mathcal{E}} \cup \plantmirror{\mathcal{E}}$ the interconnection matrix $\plant{\permute}$ is defined over the undirected graph ${\plant{\graph} + \plantmirror{\graph}}$, where $\plantmirror{\graph}$ introduces zero signals in the appropriate channels to complete the graph $\plant{\graph}$ to an undirected one. Therefore, in the following, w.\,l.\,o.\,g., undirected interconnection graphs will be considered. 

\section{Controller Structure and Closed Loop}
\label{sec:controlclosedloop}

\subsection{Interconnected Controller Structure}
\label{sec:control}
The goal of the controller synthesis is to find another interconnected system, the controller $\controrig$, such that the closed-loop system, i.e., the interconnection of the plant $\Plant$ with the controller $\controrig$, is stable and minimizes the induced $\mathcal{L}_2$-norm of the closed-loop system. 
The state-space realization of the subsystems of the controller $\controrig$ are given by 
\begin{equation}
\label{eq:dofbd}
{\controrig_i} : \begin{cases} 
\bma{@{}c@{}}{ 
	\dotxK_i \\
	u_i \\
	\hdashline
	\intercKo_i
} 
= 
\bma{@{}c@{\,\,\,\,\,\,} c@{\,\,\,\,\,\,} : c@{}}{ 
	{\AK}_i  & {\BK}_i & {B}_{\intercKi,i} \\ 
	{\CK}_i  & {\DK}_i & {C}_{\intercKi,i} \\ 
	\hdashline
	{C}_{\intercKo,i}  &  {D}_{\intercKo,i} & 0 }
\bma{@{}c@{}}{ 
	\xK_i \\
	y_i \\
	\hdashline
	\intercKi_i
}.
\end{cases} 
\end{equation}
A possible realization of the controller matrices is given in Appendix~\ref{sub:LFTMatrices_hetero} in \eqref{eq:interconn_matrices_controller_hetero}. 
As before for the plant, we define the neighboring subcontrollers of subcontroller $i$ by 
$
\cont{\mathcal{N}}_i = \{ k \vert (k,i) \in \cont{\mathcal{E}} \cup \contmirror{\mathcal{E}} \}. 
$ 
Also, the interconnection signals of the subcontroller $i$, $\intercKi_i \in \mathbb{R}^{n_{\intercKi i}}$ and $\intercKo_i \in \mathbb{R}^{n_{\intercKo i}}$, are  defined and partitioned analogously to the ones of the plant, $\intercPi_i$ and $\intercPo_i$. Also the interconnection signals of the controller, $\intercKi \in \mathbb{R}^{n_{\intercKi}}$ and $\intercKo \in \mathbb{R}^{n_{\intercKo}}$, are defined analogously to $\intercPi$ and $\intercPo$. 
With the interconnection matrix $\cont{\permute}$, the interconnection relations of the controller are defined by 
$$\intercKi =  \cont{\permute}  \intercKo.$$ 
As before for the plant, we define the interconnection graph $\cont{\graph}=\{\mathcal{N},\cont{\mathcal{E}}\}$ for the interconnected controller. Furthermore, we define the interconnection matrix $\cont{\pattern}$ as the $N \times N$-matrix capturing the topology of the controller. 
The mirror graph $\contmirror{\graph}$, as well as the mirror interconnection matrix $\contmirror{\pattern}$ are defined analogously to $\plantmirror{\graph}$ and $\plantmirror{\pattern}$ from before. 

As before for the system, we denote the decentralized part of the controller, i.e., the ensemble of all  $\controrig_i, ~\forall i=1,...,N$, by $\bd{\controrig}$. 
The interconnected controller is then given by the decentralized part $\bd{\controrig}$ and the interconnection channel as follows. 
\begin{equation}
\label{eq:interconn_control_i}
\begin{aligned}
\controrig: 
\begin{cases}
\bd{\controrig} = 
\diagIndtwo{i \in \mathcal{N}}{}{{\controrig}_i}, \\
\intercKi = \cont{\permute} \intercKo\,. 
\end{cases}
\end{aligned}
\end{equation}
Again, as for the plant, the interconnection signals, ${\controrig}_i$ and $\intercKo$, and the interconnection matrix $\cont{\permute}$ captures the topology described by the undirected graph $\cont{\graph} + \contmirror{\graph}$.

\subsection{Interconnected Closed-Loop System}

We define the closed-loop of the system $\Plant$ interconnected with the controller $\controrig$, as illustrated in the right diagram of Figure~\ref{fig:interconn_channel}, 
as $\Plantclp = \mathcal{F}_u(\Plant,\controrig) = \mathcal{F}_u(\mathcal{F}_l(\bd{\Plant},\plant{\permute}),\mathcal{F}_u(\bd{\controrig},\cont{\permute})) = \mathcal{F}_u(\mathcal{F}_l(\bd{\Plantclp},\permute))$, 
with dynamics given by 
\begin{equation}
\begin{small}
\label{eq:closed_loop_PPK_i}
\Plantclp: 
\begin{cases} 
\bd{\Plantclp} = 
\diagIndtwo{i \in \mathcal{N}}{}{\Plantclp_i}, \\
\interco 
= \permute \interci, 
\end{cases}
\end{small}
\end{equation}
with 
\begin{equation}
\Plantclp_i: 
\begin{cases} 
\small \bma{@{}c@{}}{  
	\dotxc_i  \\ \hdashline \perfo_i \\ \hdashline \interco_i  } 
\!\! 
=   
\small \bma{@{}c@{\,\,\,\,\,\,} : c@{\,\,\,\,\,\,}c@{}}{\mathcal{A}_{i} & \mathcal{B}_{1,i} & \mathcal{B}_{2,i} \\ \hdashline 
	\mathcal{C}_{1,i}  & \mathcal{D}_{11,i} & \mathcal{D}_{12,i} \\
	\mathcal{C}_{2,i}  & \mathcal{D}_{21,i} & \mathcal{D}_{22,i}}
\small \bma{@{}c@{}}{  
	\xc_i  \\ \hdashline \perfi_i \\ \hdashline \interci_i  }, ~~ \forall i \in \mathcal{N}, 
\end{cases}
\end{equation}
and where the state and interconnection signal vectors $\xc_i \in \mathbb{R}^{n_{\xc i}}$, $\interco_i \in \mathbb{R}^{n_{\interco i}}$ and $\interci_i \in \mathbb{R}^{n_{\interci i}}$ 
of the closed-loop system are defined as the stacked vectors 
of the system and the controller, as 
\begin{equation} \label{eq:defsyscon} 
\begin{aligned}
\xc_i = \bma{@{}c@{}}{ {\xP_i} \\ {\xK_i} }, \quad 
\interco_i = \bma{@{}c@{}}{ {\intercPo_i} \\ {\intercKo_i} },\quad 
\interci_i = \bma{@{}c@{}}{ {\intercPi_i} \\ {\intercKi_i} }, %
\end{aligned}
\end{equation} 
respectively. 
With 
$\xc = \concatIndtwo{i=1}{N}{\xc_i}$, $\interco = \concatIndtwo{i=1}{N}{\interco_i}$, $\interci = \concatIndtwo{i=1}{N}{\interci_i}$, 
the interconnection matrices $\pattern$ and $\permute$ are then defined for the graph of the closed-loop system, ${\graph}$, with ${\graph} = \{ \mathcal{N}, \mathcal{E}\}$, where the edge set $\mathcal{E}$ is given by 
$\mathcal{E} = \plant{\mathcal{E}} \cup \plantmirror{\mathcal{E}} \cup \cont{\mathcal{E}} \cup \contmirror{\mathcal{E}}.$  
As before in \eqref{eq:interconn_openloop}, this interconnection structure can always be achieved by introducing zero signals in the appropriate channels. 
The set of neighboring subsystems of subsystem $i$ in the closed-loop is defined analogously to the sets $\plant{\mathcal{N}}_i$ and $\cont{\mathcal{N}}_i$ for the closed loop and is thus the union of both sets, i.e., ${\mathcal{N}}_i = \plant{\mathcal{N}}_i \cup \cont{\mathcal{N}}_i$.

Note that the controller interconnection structure does not have to be the same as the plant interconnection structure, i.e., we allow for $\plant{\permute} \neq \cont{\permute}$. 
A possible realization of the closed-loop system matrices in \eqref{eq:closed_loop_PPK_i} is given in the Appendix in \eqref{eq:interconn_matrices_clp_hetero}. 

\section{Transformation to Interconnected State Space Representations}
\label{sec:PerformanceAugmentation}

\subsection{Distributed Systems with Centralized Performance}
In general, the performance channel of a given distributed plant may not be localized as it is assumed in \eqref{eq:interconn_openloop_i}. For instance, for cooperative control tasks, system-wide performance goals can be formulated, and exogenous inputs can affect coupled parts of the system. In this case, the performance channel is not localized and the system cannot readily be modeled in the form given in \eqref{eq:interconn_openloop}. 

Let us consider a distributed continuous-time LTI plant with the following dynamics
\begin{equation}
\bar{G}: 
\begin{cases}
\dotxP &= {A} \xP + \sum_{i=1}^{N} {B}_{u,i} {u}_i + {B}_{\bar{\perfi}} \bar{\perfi}\,, \\
{y}_i &= {C}_{y,i} {\xP} + D_{y\bar{\perfi},i} \bar{w} \,, \quad i=1,...,N\,, \\
\bar{\perfo} &= C_{\bar{\perfo}} \xP  + D_{\bar{\perfo}u} u ,
\end{cases}
\label{eq:overallsys_orig}
\end{equation}
with the state vector $\xP = \concatIndtwo{i=1}{N}{\xP_i} \in \mathbb{R}^{n_{\xP}}$ and the control input vector $u = \concatIndtwo{i=1}{N}{u_i} \in \mathbb{R}^{n_u}$. 
The exogenous input and performance output, which in general are not local, are given by $\bar{\perfi} \in \mathbb{R}^{n_{\bar{\perfi}}}$ and $\bar{\perfo} \in \mathbb{R}^{n_{\bar{\perfo}}}$, respectively. 

As shown in \cite{Stuerz2018a}, in order to decompose the system into local interconnected subsystems $\Plant_i$, the global performance input and output, $\bar{\perfi}$ and $\bar{\perfo}$, in \eqref{eq:overallsys_orig}, can be augmented such that to each individual subsystem a performance input and output can be assigned. This augmentation is given by 
\begin{equation}
\begin{aligned} \label{eq:perf_aug1}
\perfo &= \bar{Q}^{\frac{1}{2}} S \bar{\perfo} \quad \text{and} \quad {\perfi} = \bar{R}^{-\frac{1}{2}} T \bar{\perfi},
\end{aligned}
\end{equation}
with ${\perfo} = \concatIndtwo{i=1}{N}{{\perfo}_i}$ and ${\perfi} = \concatIndtwo{i=1}{N}{{\perfi}_i}$ and $S$ and $T$ having full rank. 
The augmentation of the related system matrices is defined as 
\begin{equation}
\label{eq:perf_aug}
\begin{aligned}
&C_{{\perfo}} = \bar{Q}^{\frac{1}{2}} S C_{\bar{\perfo}} 
\qquad \quad D_{{\perfo} u} = \bar{Q}^{\frac{1}{2}} S D_{\bar{\perfo} u}, \\
&B_{{\perfi}} = B_{\bar{\perfi}} T^\dag \bar{R}^{\frac{1}{2}} 
\qquad  D_{y {\perfi}} = D_{y \bar{\perfi}} T^\dag \bar{R}^{\frac{1}{2}}, 
\end{aligned}
\end{equation}
with $C_{{\perfo}} = \concatIndtwo{i=1}{N}{C_{\perfo,i}}$, $B_{{\perfi}}=\concatIndtwo{i=1}{N}{B_{\perfi,i}^\top}^\top$, $D_{{\perfo} u} = \concatIndtwo{i=1}{N}{D_{\perfo u, i}}$, and $D_{y {\perfi}} = \concatIndtwo{i=1}{N}{D_{y \perfi,i}^\top}^\top$.  
The matrices $\bar{Q}$ and $\bar{R}$ are weightings and are defined in the following section.

\subsection{System Norm-Invariant Transformation of the Performance Channel}
\begin{figure}
	\begin{scriptsize}
		\def\svgwidth{1\columnwidth}
		\centering\input{figures/interconn_channel_contr_2structures_.tex} 
		\centering \includegraphics[width=1\columnwidth]{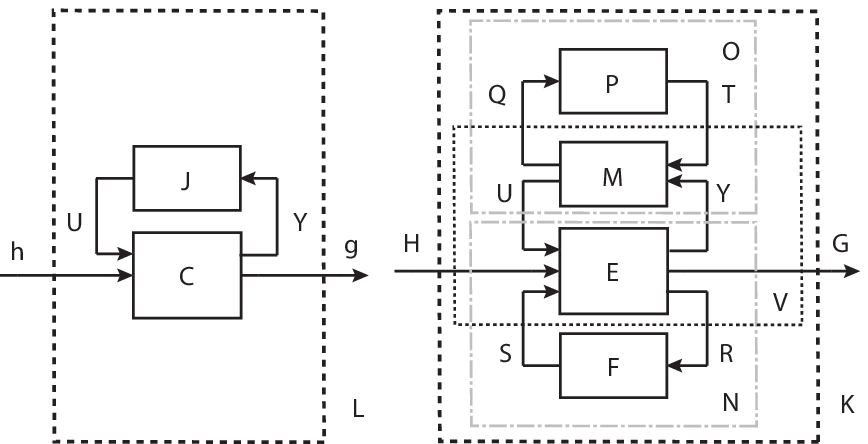} 
		\caption{\small Closed-loops of systems $\bar{\Plant}$ and $\Plant = \mathcal{F}_l(\bd{\Plant},\plant{\permute})$ with controller ${\controrig = \mathcal{F}_u(\bd{\controrig},\cont{\permute})}$. 
			The distributed performance channel $\perfo - \perfi$ of $\Plant$ allows for the interconnected structure of $\Plant = \mathcal{F}_l(\bd{\Plant},\plant{\permute})$. 
			\label{fig:interconn_channel}}
	\end{scriptsize}
\end{figure}

We define the closed-loops of the system $\bar{\Plant}$ in \eqref{eq:overallsys_orig} and $\Plant$ in \eqref{eq:interconn_openloop} 
in interconnection with a controller $\controrig$ as $\bar{\Plantclp} = \mathcal{F}_u(\bar{\Plant},\controrig)$ and $\Plantclp = \mathcal{F}_u(\Plant,\controrig)$, respectively. The closed-loops are illustrated in Figure~\ref{fig:interconn_channel}. 

As control objective, the $\mathcal{H}_\infty$-norm of the closed-loop transfer function from $\bar{w}$ to $\bar{z}$ of the system $\bar{G}$ 
under the controller $\controrig$, i.e. $\| \bar{\Plantclp} \|_{\mathcal{H}_\infty}$, is to be minimized. 
For a scalable synthesis of $\controrig$, the goal is to exploit the structure of $\Plant$ 
and thus to minimize $\| {\Plantclp} \|_{\mathcal{H}_\infty}$. 
We show that under the same controller $\controrig$ this norm is equal to the norm of ${\Plantclp}$, 
i.e., $\| \bar{\Plantclp} \|_{\mathcal{H}_\infty} = \| \Plantclp \|_{\mathcal{H}_\infty}$. 
Let us assume that the full rank matrices $\bar{Q}$ and $\bar{R}$ are chosen as 
\begin{equation}
\label{eq:Qbar}
\begin{array}{cll}
&\bar{Q} = {S^\dag}^\trans   S^\dag + M_{Q} ,   \quad\quad  &\bar{R} = T T^\trans + M_{R}, 
\end{array}
\end{equation}
with $M_Q = M_Q^\trans$,   $M_R = M_R^\trans$, $S^\trans M_{Q} S = 0$, and  $T^\dag M_{R} {T^\dag}^\trans = 0$. 
Then, the equality of $\| \bar{\Plantclp} \|_{\mathcal{H}_\infty}$ and $\| \Plantclp \|_{\mathcal{H}_\infty}$ is stated in the following. 
\begin{mythe}
	\label{the:normequivalence}
	Given $\Plantclp$  and $\mathcal{\bar{G}}$, with the transformation in \eqref{eq:perf_aug} and the weightings in \eqref{eq:Qbar}, it holds that 
	\begin{equation*}
	\| \bar{\Plantclp} \|_{\mathcal{H}_\infty} = \| \Plantclp \|_{\mathcal{H}_\infty}.
	\end{equation*}
\end{mythe}
\begin{IEEEproof}
	Similarly as in Lemma~9 in \cite{Massioni2009}, for a transformation of a system $\bar{\Plantclp}$ to ${\Plantclp}$ with $\Plantclp = T_l \bar{\Plantclp} T_r^\dag$, the following performance bounds can be proven    
	\begin{equation}
	\label{eq:perf_bounds}
	\begin{aligned}
	\frac{\sigma_{\mathrm{min}}(T_r)}{\sigma_{\mathrm{max}}(T_l)} \| {\Plantclp} \|_{\mathcal{H}_\infty} 
	&\leq  
	\| \bar{\Plantclp} \|_{\mathcal{H}_\infty} \leq 
	\frac{\sigma_{\mathrm{max}}(T_r)}{\sigma_{\mathrm{min}}(T_l)}  \| {\Plantclp} \|_{\mathcal{H}_\infty}. \\ 
	\end{aligned}
	\end{equation}
	With $T_l := {\left( S^{\dag T}   S^\dag + M_{Q} \right)}^{\frac{1}{2}} S$ and $T_r := {\left( T  T^T + M_{R} \right)}^{-\frac{1}{2}} T$, 
	we need to show that the transformation matrices $T_l$ and $T_r$ are semi-orthogonal, which is a generalization of orthogonality for rectangular matrices, i.e., $T_r^\trans T_r = I$ and $T_l T_l^\trans = I$. 
	Semi-orthogonal $m \times n$ or $n \times m$-matrices have $m$ singular values of 1, if $m \leq n$. 
	Then, the bounds in \eqref{eq:perf_bounds} are tight since $\sigma_{max} = \sigma_{min} = 1$ and so the $\mathcal{H}_\infty$-norm is not changed under the system transformation. 
	To show that $T_r$ is semi-orthogonal, 
	we see that 
	\begin{equation*}
	\begin{aligned}
	T_r^\trans T_r &= \left(  ( T T^\trans + M_R )^{-\frac{1}{2}} T \right)^\trans  \left(  ( T T^\trans + M_R )^{-\frac{1}{2}} T \right) \\
	&= T^\trans  \left( ( T T^\trans + M_R )^{-\frac{1}{2}} \right)^\trans    ( T T^\trans + M_R )^{-\frac{1}{2}} T \\
	&= T^\trans  ( T T^\trans + M_R )^{-1}  T 
	= I ,
	\end{aligned}
	\end{equation*}
	which holds because of $M_R = M_R^\trans$. Showing that $T_l^\trans T_l = I$ follows along the same lines. 
\end{IEEEproof}
The following proposition suggests that the augmentation of the performance channel is also applicable for an $\mathcal{H}_2$-based controller synthesis. 
\begin{myprop}
	Given $\Plantclp$ and $\bar{\Plantclp}$, it holds 
	$$\| \Plantclp \|_{\mathcal{H}_2} = \|\bar{\Plantclp} \|_{\mathcal{H}_2}.$$ 
\end{myprop}
\begin{IEEEproof}
	As shown in \cite{Massioni2009}, the $\mathcal{H}_2$-norm is also unitary-invariant, and therefore the same proof as in Theorem~\ref{the:normequivalence} can be applied to show that $\| \Plantclp \|_{\mathcal{H}_2} = \| \bar{\Plantclp} \|_{\mathcal{H}_2}$. 
\end{IEEEproof}

This performance input-output-transformation in \eqref{eq:perf_aug1} and \eqref{eq:perf_aug} with \eqref{eq:Qbar} leads to the system 
\begin{equation}
\Plant: 
\begin{cases}
\dotxP &= {A} \xP + \sum_{i=1}^{N} {B}_{u,i} {u}_i + \sum_{i=1}^{N} {B}_{\perfi, i} {\perfi}_i\,, \\ 
{y}_i &= {C}_{y,i} \xP + D_{y\perfi, i} \perfi \,, \quad i=1,...,N\,, \\
\perfo_i &= C_{\perfo, i} \xP  + D_{\perfo u,i} u ,
\end{cases}
\label{eq:overallsys}
\end{equation}
which has local control and performance inputs and outputs, $u_i$, $y_i$, $\perfi_i$, and $\perfo_i$, respectively, and can thus readily be modeled as the interconnected system in \eqref{eq:interconn_openloop}.

\section{Decomposed Controller Synthesis Equations}
\label{sec:DecomposedEquations}

In the following, we will consider methods from robust and gain-scheduled controller synthesis, where the interconnection plays the role of the uncertainty. The block-diagonal entries in the system matrices are dealt with in a decomposed way and the off-block-diagonal entries modeled through the interconnection channel need to be accounted for in multiplier conditions that are coupled over the subsystems. 
We present a decomposed synthesis for the interconnected controller $\controrig$ based on the system representation in \eqref{eq:closed_loop_PPK_i}. 

\subsection{Centralized Full Block S-Procedure for Interconnected Systems}
In the following, we will use the full block S-procedure for the controller synthesis, and therefore we briefly review it here. 
\begin{mythe}[\cite{Scherer2001}: Full Block S-Procedure]
	\label{the:sproc}
	Given the stable continuous-time LTI system 
	\begin{equation} \label{eq:closed_loop}
	\small \bma{@{}c@{}}{  
		\dotxc \\ \hdashline \perfo \\  \interco} =
	\small \bma{@{}c@{\,\,\,\,\,\,} : c@{\,\,\,\,\,\,}c@{}}{\mathcal{A} & \mathcal{B}_1 & \mathcal{B}_2 \\ \hdashline 
		\mathcal{C}_1  & \mathcal{D}_{11} & \mathcal{D}_{12} \\
		\mathcal{C}_2  & \mathcal{D}_{21} & \mathcal{D}_{22}}
	\!\! \!\!
	\small \bma{@{}c@{}}{  
		\xc  \\ \hdashline \perfi \\  \interci}, \quad \interco = \permute \interci,
	\end{equation}
	then the system has an $\mathcal{L}_2$-gain from $\perfi$ to $\perfo$ smaller than $\gamma$ if and only if there exist variables $\mathcal{X} = \mathcal{X}^\trans > 0$, $R= R^\trans$, $Q= Q^\trans$ and $S$ of appropriate dimensions such that 
	\begin{align}
	\label{eq:primalLMIs_mult}
	\bma{@{}c@{}}{ 
		\star
	}^\top 
	\small \bma{@{}c@{\,\,\,\,\,} c@{}}{ 
		Q   & S \\
		S^\trans & R
	}^\trans
	\small \bma{@{}c@{}}{ 
		\permute \\
		I } 
	& > 0  , \\
	\bma{@{}c@{}}{ 
		\star
	}^\trans 
	\small \bma{@{}c@{\,\,\,\,\,} c@{\,\,\,\,\,} : c@{\,\,\,\,\,}  c@{\,\,\,\,\,} : c@{}  c@{}}{ 
		0 & \mathcal{X} & 0 & 0 & 0 & 0 \\
		\mathcal{X} & 0 & 0 & 0 & 0 & 0 \\
		\hdashline
		0 & 0 & - \gamma I & 0 & 0 & 0 \\
		0 & 0 & 0 & \frac{1}{\gamma} I & 0 & 0 \\
		\hdashline
		0 & 0 & 0 & 0 &  Q & S \\
		0 & 0 & 0 & 0 & S^\trans & R
	} 
	\small \bma{@{}c@{\,\,\,\,\,\,} c@{\,\,\,\,\,\,}  c@{}}{ 
		I & 0 & 0 \\
		\mathcal{A} & \mathcal{B}_1 & \mathcal{B}_2 \\
		\hdashline
		0 & I & 0 \\
		\mathcal{C}_1 & \mathcal{D}_{11} & \mathcal{D}_{12} \\
		\hdashline
		0 & 0 & I \\
		\mathcal{C}_2 & \mathcal{D}_{21} & \mathcal{D}_{22}
	} 
	&< 0 .
	\label{eq:primalLMIs}
	\end{align}
\end{mythe}
\vspace{0.2cm}
This theorem can be directly applied to the system formulation in \eqref{eq:closed_loop_PPK_i}. 
We show in the following proposition how the conditions in Theorem~\ref{the:sproc} 
can then be decomposed into conditions of the size of the subsystems by appropriate structural (block-diagonal) assumptions on the multipliers $Q$, $R$ and $S$ and on the Lyapunov matrix $\mathcal{X}$. 
We focus on the case of heterogeneous systems, for which we propose a distributed synthesis in Section \ref{sec:DistributedDesign}. 
In Section~\ref{sec:SpecialCases}, we present two special cases of systems 
for which the decomposed synthesis equations can be significantly simplified under further structural assumptions.

\subsection{Decomposed Controller Synthesis for Heterogeneous Systems}

\begin{myprop}
	\label{prop:smallLMIs_hetero}
	\textit{(Decomposed Full-Block S-Procedure for Heterogeneous Systems:)}  
	Consider a heterogeneous system $\Plant = \LFTl(\bd{\Plant},\plant{\permute})$ given in \eqref{eq:interconn_openloop} 
	Then there exists a controller $\controrig$ 
	as in \eqref{eq:dofbd} such that ${\Plantclp} = \LFTu(\Plant,\controrig)  = \LFTu(\bd{\Plantclp},\permute)$ in \eqref{eq:closed_loop_PPK_i} 
	is stable and has an 
	$\mathcal{L}_2$-gain 
	less than $\gamma$, if there exist $\mathcal{X}_i = \mathcal{X}_i^\trans > 0$ and 
	${R}_{ik}= {R}_{ik}^\trans$, ${Q}_{ik}= {Q}_{ik}^\trans$ and ${S}_{ik}$,  $\forall (i,k) \in \mathcal{E}$, 	
	such that
	\begin{equation}
	\begin{aligned}
	\label{eq:primalLMIs_small_mult_hetero}
	\small \bma{@{}c@{}}{ \star }^\top 
	\small \bma{@{}c@{\,\,\,\,\,\,} c@{}}{ 
		\hat{Q}_{ik}  & \hat{S}_{ik} \\
		\hat{S}_{ik}^\trans  & \hat{R}_{ik}} 
	\small \bma{@{}c@{}}{ 
		\hat{\permute}_{ik} \\ 
		I_{(n_{\interci_{ik}}+n_{\interci_{ki}})}  } &>0, \\[0.2cm]
	\forall (i,k) \in \mathcal{E}. & 
\end{aligned}
\end{equation}
\begin{align}
	\bma{@{}c@{}}{ 
		\star
	}^\trans 
	\small \bma{@{}c@{\,\,\,\,\,} c@{\,\,\,\,\,} : c@{\,\,\,\,\,}  c@{\,\,\,\,\,} : c@{}  c@{}}{ 
		0 & \mathcal{X}_i  & 0 & 0 & 0 & 0 \\
		\mathcal{X}_i  & 0 & 0 & 0 & 0 & 0 \\
		\hdashline
		0 & 0 &  - \gamma I & 0 & 0 & 0 \\
		0 & 0 & 0 & \frac{1}{\gamma} I & 0 & 0 \\
		\hdashline
		0 & 0 & 0 & 0 & \smallQ{Q}_i  & \smallQ{S}_i  \\
		0 & 0 & 0 & 0 &  \smallQ{S}_i^\trans  & \smallQ{R}_i 
	} 
	\small \bma{@{}c@{\,\,\,\,\,\,} c@{\,\,\,\,\,\,}  c@{}}{ 
		I & 0 & 0 \\
		\mathcal{A}_i & \mathcal{B}_{1,i} & \mathcal{B}_{2,i} \\
		\hdashline
		0 & I & 0 \\
		\mathcal{C}_{1,i} & \mathcal{D}_{11,i} & \mathcal{D}_{12,i} \\
		\hdashline
		0 & 0 & I \\
		\mathcal{C}_{2,i} & \mathcal{D}_{21,i} & \mathcal{D}_{22,i}
	} 
	&< 0 , \notag\\
	\forall i \in \{1,...,N\}. &
	\label{eq:primalLMIs_small_hetero}
	\end{align}
		with 
	\begin{displaymath}	\smallQ{R}_i= \diagIndtwo{k \in \mathcal{N}_i}{}{{R}_{ik}}, ~
		\smallQ{Q}_i= \diagIndtwo{k \in \mathcal{N}_i}{}{{Q}_{ik}}, ~
		\smallQ{S}_i= \diagIndtwo{k \in \mathcal{N}_i}{}{{S}_{ik}}, 
		\end{displaymath}	
	\begin{displaymath}
	\hat{R}_{ik}=\mathrm{diag}({R}_{ik},R_{ki}), ~~
		\hat{Q}_{ik}=\mathrm{diag}({Q}_{ik},{Q}_{ki}), ~~
		\hat{S}_{ik}=\mathrm{diag}({S}_{ik},{S}_{ki}),
	\end{displaymath} 
	and with 
	$$\hat{\permute}_{ik} = \small \bma{@{}c@{\,\,\,\,\,\,} c@{}}{ 
		0 & \permute_{ik}  \\
		\permute_{ki}  & 0} .$$
\end{myprop}
\begin{IEEEproof}
	When applying Theorem~\ref{the:sproc} to the interconnected system in \eqref{eq:closed_loop_PPK_i} with the structured Lyapunov matrix $\mathcal{X} = \kronecker{I_N}{{\mathcal{X}_i}}$ and the structured multipliers ${R}=\diagIndtwo{i \in \mathcal{N}}{}{\smallQ{R}_i}$, ${Q}=\diagIndtwo{i \in \mathcal{N}}{}{\smallQ{Q}_i}$, and 
	${S}=\diagIndtwo{i \in \mathcal{N}}{}{\smallQ{S}_i}$, 
	then the matrices in the nominal condition \eqref{eq:primalLMIs} are composed of only block-diagonal matrices and therefore completely decompose into one condition per subsystem as in \eqref{eq:primalLMIs_small_hetero}.

With ${\graph}$ being an undirected graph, and with the structured multipliers 
$\smallQ{R}_i= \diagIndtwo{k \in \mathcal{N}_i}{}{{R}_{ik}}$, $\smallQ{Q}_i= \diagIndtwo{k \in \mathcal{N}_i}{}{{Q}_{ik}}$, and $\smallQ{S}_i= \diagIndtwo{k \in \mathcal{N}_i}{}{{S}_{ik}}$, 
the multiplier condition in \eqref{eq:primalLMIs_mult} can be transformed into a block-diagonal matrix with the conditions of \eqref{eq:primalLMIs_small_mult_hetero} on its diagonal blocks. 
To see this, let us consider the interconnection channel $\interci = \permute \interco$ as defined in \eqref{eq:closed_loop_PPK_i}. 
Then, we can always find a permutation matrix $T$, such that the entries of the signals $\interco$ and $\interci$ are reordered such that those corresponding to the same edge are consecutive, i.e., $\bar{\interci} = T \interci$ with $\bar{\interci} = [..., {\interci_{ik}}^\trans, \,\, {\interci_{ki}}^\trans, ...]^\trans$, and $\bar{\interco} = T \interco$ with $\bar{\interco} = [..., {\interco_{ik}}^\trans, \,\, {\interco_{ki}}^\trans, ...]^\trans$. 
The similarity transformation of the multiplier condition with $T$ leads to 
\begin{equation}\label{eq:trafoT}
\small \bma{@{}c@{}}{ \star }^T 
\underbrace{\small \bma{@{}c@{\,\,\,\,\,\,} c@{}}{ 
	{T}  &  \\
	& T} 
\small \bma{@{}c@{\,\,\,\,\,\,} c@{}}{ 
	{Q}  & {S} \\
	{S}^\trans  & {R}} 
\small \bma{@{}c@{\,\,\,\,\,\,} c@{}}{ 
	{T}^{-1}  &  \\
	& {T}^{-1}} }_{\small \bma{@{}c@{\,\,\,\,\,\,} c@{}}{ 
	\bar{Q}  & \bar{S} \\
	\bar{S}^\trans  & \bar{R}} }
\underbrace{\small \bma{@{}c@{\,\,\,\,\,\,} c@{}}{ 
	{T}  & \\
	& {T}} 
\small \bma{@{}c@{}}{ 
	{\permute} \\
	I_{n_{\interci}}  }
	{T}^{-1}}_{\small \bma{@{}c@{}}{ 
		\bar{\permute} \\
		I_{n_{\interci}}  }}
  >0,
\end{equation}
involving the multiplier transformations 
$\bar{Q} = T Q T^{-1}$, $\bar{S}= T S T^{-1}$ and $\bar{R}= T R T^{-1}$. 
As the multipliers $Q$, $R$ and $S$ are block-diagonal, this transformation results in block-diagonal multipliers again 
with the same reordering of the blocks on the diagonal as for the interconnection signals. 
This completes the proof that $\controrig$ stabilizes $\Plantclp = \LFTu(\Plant,\controrig)$ and leads to an 
	$\mathcal{H}_\infty$-norm less than $\gamma$.  
\end{IEEEproof}
The transformation in \eqref{eq:trafoT} is illustrated in Example~\ref{exp:trafo_blkdiag} in Section~\ref{sec:NumericalExample}. 
\begin{mycor} \label{cor:decSproc_GbarG}
	The controller $\controrig$ in Proposition~\ref{prop:smallLMIs_hetero} stabilizes $\bar{\Plantclp}$ and leads to a performance bound of less than $\gamma$ for $\bar{\Plantclp}$. 
\end{mycor}		
\begin{IEEEproof}
	Note that from $G$ to $\bar{G}$, only the performance channel is transformed, and therefore stability of $\Plantclp$, which is guaranteed by Proposition~\ref{prop:smallLMIs_hetero}, implies stability of $\bar{\Plantclp}=\LFTu(\Plant,\controrig)$. Furthermore, it has been shown in Theorem~\ref{the:normequivalence} that the $\mathcal{H}_\infty$-norm is invariant under the transformation of the performance channel, and thus the performance bound $\gamma$ on $\Plantclp$ also holds for $\bar{\Plantclp}$. 
\end{IEEEproof}

\begin{myrem} 
	In the case of ideal interconnections, where $\permute_{ik}$ are identities, then \eqref{eq:primalLMIs_small_mult_hetero} results in the simplified conditions 
	\begin{equation}\label{eq:LMI_ik} 
	\begin{aligned}
	 \small \bma{@{}c@{\,\,\,\,\,\,\,\,\,\,\,\,}c@{}}{
		Q_{ik}  + R_{ki}&  S_{ik} + S^\trans_{ki}  \\[0.1cm]
		S_{ki} + S^\trans_{ik}  &  Q_{ki}  + R_{ik} \\
	} > 0,  
	~~~ \forall (i,k) \in \mathcal{E}. 
	\end{aligned}
	\end{equation} 
\end{myrem}

\section{Distributed Controller Synthesis for Heterogeneous Systems}
\label{sec:DistributedDesign}

The controller synthesis is based on the nominal conditions in \eqref{eq:primalLMIs_small_hetero} and on the multiplier conditions in \eqref{eq:primalLMIs_small_mult_hetero}. 
While the nominal conditions are completely decomposed into small conditions for each subsystem, the decomposed multiplier conditions in Proposition~\ref{prop:smallLMIs_hetero} introduce pairwise coupling between interconnected neighboring subsystems. 
Therefore, a decentralized controller synthesis is not possible and we seek a distributed design method in the following. 
Our approach is based on a variant of the consensus ADMM \cite{Boyd2010}, where only nearest neighbor communication is required. This variant was introduced in \cite{Mateos2010} and extended in \cite{Chang2015, Banjac2018}.

\subsection{Decomposed Controller Synthesis Problem}

We define the global variable vector $\gl$ containing the set of global controller gains 
$\{\AK, \BK, \CK, \DK\}$ of $\controrig$ in \eqref{eq:interconn_control_i} with all local controller gains in \eqref{eq:dofbd}, 
the global (block-diagonal) Lyapunov matrix $\mathcal{X}$, 
all multipliers ${Q}_{ik},{R}_{ik},{S}_{ik}$ 
from Proposition~\ref{prop:smallLMIs_hetero}, 
and $\gammainv = \gamma^{-\frac{1}{2}}$, with $\gamma$ being the performance bound in Theorem~\ref{the:sproc}. 
The global synthesis problem can then be formulated as 
\begin{equation} \label{eq:globalsyncentral}
\begin{aligned}
\min_{\gl} & ~~f(\gl) + g(\gl),
\end{aligned}
\end{equation}
with $f(\cdot)$ and $g(\cdot)$ being defined as 
\begin{equation}
\begin{aligned}
f(\gl) &= - \gammainv,  \\ 
g(\gl) &= \mathcal{I}_{\eqref{eq:primalLMIs_mult}}(\gl) + \mathcal{I}_{\eqref{eq:primalLMIs}}(\gl
).
\end{aligned}
\end{equation}
Herein, $\mathcal{I}_{(r)}(z)$ denotes the indicator function of $z$ satisfying the conditions in $(r)$, i.e., 
\begin{equation}
\begin{aligned}
&\mathcal{I}_{(r)}(z) := 
\begin{cases} 
0 ~\mathrm{if} ~ z ~\mathrm{satisfies} ~ (r), \\
\infty ~\mathrm{otherwise}. 
\end{cases} 
\end{aligned}
\end{equation} 
In order to decompose the global synthesis problem, we introduce the set of local variables $\y_i$ for all subsystems $i \in \mathcal{N}$, which contain copies of all variables of the global variable vector $\gl$ that are relevant to the respective subsystem $i$. 
The local variable vector $\y_i$ thus contains the local controller gains  
\begin{equation}\label{eq:Ki}
\begin{aligned}
\{ &\AK_i,\BK_i,\CK_i,\DK_i, \\ &\AK_{ik},\BK_{ik},\CK_{ik},\DK_{ik},~\forall k \in \cont{\mathcal{N}}_i \}
\end{aligned}
\end{equation}
of $\controrig_i$ in \eqref{eq:dofbd}, the local Lyapunov matrices $\mathcal{X}_i$  of $\mathcal{X} = \diagIndtwo{i=1}{N}{\mathcal{X}_i}$, 
the local copies $\gammainv_i$ of $\gammainv$, and the local copies of the multipliers involved in the local synthesis problem \eqref{eq:primalLMIs_small_mult_hetero}, \eqref{eq:primalLMIs_small_hetero} of subsystem $i$. 
In order to ensure consistency over local copies of variables by different interconnected subsystems $i$ and $k$ corresponding to the same parts of the global variable $\gl$, we further define the selection matrices ${E}_{ik}$ and $E_{ki}$ and formulate 
the following local consensus constraint 
$${E}_{ik} \, \y_i = {E}_{ki} \, \y_k, ~~ \forall (i,k) \in \mathcal{E}.$$ 
We further define the dual multipliers of subsystem $i$ corresponding to the consensus constraints for $\y_i$ 
as $\lambda_i$. 
The decomposed controller synthesis problem is now expressed as 
\begin{equation} \label{eq:decomposedsyncentral}
\begin{aligned}
\min_{\y_i}
& ~~ \sum_{i=1}^{N} \left( f_i(\y_i) + g_i(\y_i)
\right), \\
\mathrm{s.t.} ~~ & ~~ {E}_{ik} \, \y_i = {E}_{ki} \, \y_k, ~~ \forall (i,k) \in \mathcal{E}. 
\end{aligned}
\end{equation}
with $f_i(\cdot)$ and $g_i(\cdot)$ being defined as 
\begin{equation}
\begin{aligned}
f_i(\y_i) &= - \frac{1}{N} \gammainv_i, \\ 
g_i(\y_i) &= 
\mathcal{I}_{\eqref{eq:primalLMIs_small_mult_hetero}}(\y_i) +  \mathcal{I}_{\eqref{eq:primalLMIs_small_hetero}}(\y_i).
\end{aligned}
\end{equation}

\subsection{Distributed Synthesis Without Global Coordination}

We present the distributed consensus ADMM algorithm in Algorithm~\ref{alg:ADMM} with only nearest neighbor communication to solve the interconnected controller synthesis problem of Proposition~\ref{prop:smallLMIs_hetero}. The communication during the synthesis is defined over the graph ${\graph} = \{ \mathcal{N}, \mathcal{E}\}$, where the edge set $\mathcal{E}$ is given by 
$\mathcal{E} = \plant{\mathcal{E}} \cup \plantmirror{\mathcal{E}} \cup \cont{\mathcal{E}} \cup \contmirror{\mathcal{E}}$,  as introduced in Section~\ref{sec:controlclosedloop}.B.

\begin{algorithm}[t!]
	\SetAlgoLined
		\caption{Consensus ADMM}\label{alg:ADMM} 
			\KwIn{Parameter $\rho > 0$,  
			local subsystems $G_i$, $\forall i = 1,...,N$,  
			interconnections $\plant{\permute}_{ik}$ $\forall k \in \plant{\mathcal{N}}_i$, \\ 
			$\cont{\permute}_{ik}$ $\forall k \in \cont{\mathcal{N}}_i$, ~~$\forall i = 1,...,N,$ , \\ 
			initial value $\y_i^{(0)}$, $\forall i \in \mathcal{N}$} 
			\textbf{Initialization}: Set $\kappa \gets 0$, $\lambda_i^{(0)} \gets 0$\; 
			\While{\textit{not converged}}{
			Communicate $E_{ik} \, \y_i^{(\kappa)}$ to neighboring nodes $k$ in $\mathcal{N}_i$\;
			$\lambda_i^{(\kappa + 1)} \gets \lambda_i^{(\kappa)} + \rho \sum_{k \in \mathcal{N}_i}(T_{ik} \, \y_i^{(\kappa)} - T_{ki} \, \y_k^{(\kappa)})$\;
			\begin{equation*} 
			\begin{aligned} 
			\vspace{-0.2cm} 
			\hspace{-0.2cm} 
			\y_i^{(\kappa + 1)} \gets  & ~\underset{\y_i}{\mathrm{argmin}} \Bigg\{ f_i(\y_i) + g_i(\y_i) + 
			\y_i^\top \lambda_i^{(\kappa+1)} 
			\\ & + \rho \mathlarger{\mathlarger{\sum}}_{k \in \mathcal{N}_i} \norm{ T_{ik} \, \y_i - \frac{T_{ik} \, \y_i^{(\kappa)}+ T_{ki} \, \y_k^{(\kappa)}}{2} }_2^2 \Bigg\}; 
			\end{aligned} 
			\end{equation*}
			$\kappa \gets \kappa+1$\;
			}
			\KwOut{
			local controller gains in \eqref{eq:Ki}, 
			bounds $\gamma_i = \gamma$} 
	\end{algorithm}
	
In steps 4 and 5, the matrices $T_{ik}$ and $T_{ki}$ select the elements over which a consensus should be reached and place them where they occur in $\y_i$. 
A similar derivation of the steps in Algorithm~\ref{alg:ADMM} can be found in \cite{Mateos2010} and is for convenience given in Appendix~\ref{sec:derivationADMM}. 

The following primal and dual residuals \cite{Boyd2010}, can be considered as convergence criteria, 
\begin{equation} \label{eq:residuals}
\begin{aligned}
r^{(\iter + 1)} &=  \concatIndtwo{i = 1}{N}{ \concatIndtwo{k \in \mathcal{N}_i}{}{\frac{1}{2} r_{ik}^{(\iter+1)} } }, \\
d^{(\iter + 1)} &=  \concatIndtwo{i = 1}{N}{ \concatIndtwo{k \in \mathcal{N}_i}{}{\frac{1}{2}  d_{ik}^{(\iter+1)}  } }, 
\end{aligned}
\end{equation}
respectively, with
\begin{equation}
\begin{aligned}
r_{ik}^{(\iter+1)} &= E_{ik} \y_i^{(\iter + 1)} - E_{ki} \y_k^{(\iter + 1)}, \\
d_{ik}^{(\iter+1)} &= E_{ik} ( \y_i^{(\iter + 1)} - \y_i^{(\iter)} )   
+   E_{ki} ( \y_k^{(\iter + 1)} - \y_k^{(\iter)} ). 
\end{aligned}
\end{equation}
The derivation of these residuals is given in \eqref{eq:derivation_residuals} in Appendix~\ref{sub:LFTMatrices_hetero}.

Note that for determining convergence, the primal and dual residuals, $r_i$ and $d_i$, can be computed locally. Therefore, some higher-level communication protocol of low communication frequency is required to detect when convergence among all subsystems is reached.

\begin{myrem}\label{rem:SFOF}
In the case of state feedback control, variable substitutions in the synthesis equations lead to a convex problem (LMIs). In the case of dynamic output feedback, a variable substitution leads to a bilinear program (BMIs), which can be dealt with by iteratively solving two LMIs. This holds for the global problem formulation in (18), (19), and importantly, it also holds for the decomposed problem formulation in \eqref{eq:primalLMIs_small_mult_hetero}, \eqref{eq:primalLMIs_small_hetero}. 
Therefore, in the case where an interconnected state feedback controller is to be designed, the synthesis problem in \eqref{eq:primalLMIs_small_mult_hetero}, \eqref{eq:primalLMIs_small_hetero} is convex and the convergence results in \cite{Boyd2010} hold for the distributed synthesis in Algorithm~\ref{alg:ADMM}. 
If the controller to be synthesized is a dynamic output feedback controller, the non-convex (bilinear) decomposed synthesis equations in \eqref{eq:primalLMIs_small_mult_hetero}, \eqref{eq:primalLMIs_small_hetero} could be solved iteratively in step 6 in Algorithm~\ref{alg:ADMM}. No convergence guarantee for the ADMM iterations can be given in this case.  
In the case of distributed output feedback control design, further numerical techniques could be investigated in order to reduce the number of iterations to convergence of the proposed ADMM scheme, such as warm-starting with solutions of the previous ADMM iteration or early termination, such as in \cite{Darup2019} for real-time ADMM. 
\end{myrem}

\begin{myrem} 
Note that in addition to imposing the block-diagonal structure on the multipliers, one could consider further restricting them to DG scalings or diagonal multipliers, which would reduce the dimension of the consensus variables, and therefore both the dimension of the communicated signals as well as the computation and convergence time could possibly be reduced. 
Since this would introduce more conservatism, we chose to allow for (block-diagonally structured) full blocks. 
\end{myrem}

\section{Decomposed Synthesis for Special Classes of Interconnected Systems} 
\label{sec:SpecialCases}

Depending on the degree of homogeneity of the subsystems, it can be beneficial to introduce a special class of interconnected systems. First, we review the case of homogeneous systems and then introduce a new class of systems, referred to as $\aalpha$-$\intercN$-heterogeneous systems. We show how the interconnected system model in \eqref{eq:interconn_openloop_i} can be transformed into a more compact model for this system classification thereby leading to more compact controller synthesis formulations.

\subsection{Homogeneous Systems}
\begin{figure}
	\begin{center}
	\input{figures/interconn_channel_trafo.tex} 
	\includegraphics[width=0.8\columnwidth]{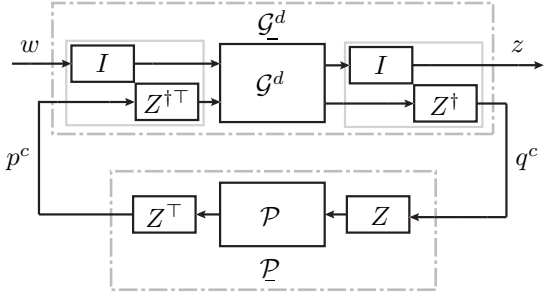} 
	\caption{Transformation of the interconnection channel.  \label{fig:interconn_trafo}}
	\end{center}
\end{figure}

\begin{mydef}[Homogeneous system]
	Let $M$ represent all overall system matrices $A$, $B_u$, $C_y$, $D_{y \perfi}$, $C_{\perfo}$, $D_{\perfo u}$, $D_{\perfo \perfi}$, which are the stacked subsystem matrices from \eqref{eq:overallsys}, i.e., ${B_u = \concatIndtwo{i=1}{N}{B_{ui}^\trans}^\trans}$, ${C_y= \concatIndtwo{i=1}{N}{C_{yi}}}$ and the other matrices are defined analogously. 
	Then, we define a homogeneous system if its system matrices can be written as 
	\begin{equation}
	\label{eq:M_homo}
	M = \underbrace{ \kronecker{I_N}{ {{M}_{ ii}}}  }_{\bd{M}} + 
	\underbrace{ \kronecker{ \plant{\pattern} }{{M}_{ik}} }_{\obd{M}},  
	\end{equation}
	with $\plant{\pattern}$ as defined in Section~\ref{subsec:Graphstructure}. 
	This means that for a homogeneous system all its local subsystem matrices $M_{ii}, ~ \forall i \in \mathcal{N}$ are identical and all its interconnection subsystem matrices $M_{ik}, ~ \forall (i,k) \in \plant{\mathcal{E}}$ are identical. 
\end{mydef}
\begin{myprop}
If the controller $\controrig$ is also chosen as a homogeneous system, i.e., such that \eqref{eq:M_homo} holds with $M$ representing the controller matrices $\AK$, $\BK$, $\CK$ and $\DK$, and if $\cont{\pattern} = \plant{\pattern}$, 
then, the interconnected closed loop system in \eqref{eq:closed_loop_PPK_i} can be transformed to a representation where 
the interconnection matrix takes the form 
\begin{equation} \label{eq:P_homo}
\barbelow{\permute} 
= 
\kronecker{{\pattern}}{I_{n_{\interci}}}, 
\end{equation}
with ${\pattern} = \plant{\pattern} = \cont{\pattern}$, 
and the system matrices of \eqref{eq:interconn_openloop_i} are transformed such that the closed-loop is not changed. 
\end{myprop}
\begin{IEEEproof}
We define the transformations of the interconnection channel, ${\permute}$, and of the decentralized system part, $\bd{\Plantclp}$, as 
\begin{equation} \label{eq:Ztransform}
\begin{aligned}
\barbelow{{\permute}} &= Z^\trans {\permute} Z, \\  
\barbelow{\bd{\Plantclp}} &= \diagonal \left( I, Z^{\dag} \right)  \bd{\Plantclp}  \diagonal \left( I, Z^{\dag \trans} \right), 
\end{aligned}
\end{equation}
respectively. 
The transformation matrix $Z$ is defined as the $\vert \mathcal{E} \vert \times N$ matrix of all zeros except for ones in the entries corresponding to an interconnection between an edge and a system. With this transformation, the interconnection topology is captured in ${\permute} = \kronecker{{\pattern}}{I_{n_\interci}}$ and the interconnections  are summarized into one channel per subsystem. This transformation is shown in Figure~\ref{fig:interconn_trafo}, and is illustrated in Example~\ref{exp:trafo_homo}. A possible realization of the transformed system matrices of $\barbelow{\bd{\Plantclp}}$ is given in the Appendix in \eqref{eq:interconn_matrices_homo}. 
\end{IEEEproof}

This model of interconnected systems has been used in \cite{Massioni2009} and in \cite{Hoffmann2015}. 
In this representation, the controller synthesis equations can be simplified as follows.

\subsection{Decomposed Controller Synthesis for Homogeneous Systems} 

\begin{myprop}
	\label{prop:smallLMIs_homo}
	\textit{(Decomposed Full-Block S-Procedure for Homogeneous Systems \cite{Massioni2009}:)}  
	Let us consider a homogeneous system ${\Plant} \! =\! \LFTl(\bd{{\Plant}},\plant{\permute})$, which is given in \eqref{eq:interconn_openloop} with $\plant{\permute} = \kronecker{\plant{\pattern}}{I_{n_{\intercPi}}}$ structured as in \eqref{eq:P_homo}, assumed to be normal, and the system matrices structured as in \eqref{eq:M_homo}.  
	Then there exists a controller $\controrig$ 
	as in \eqref{eq:dofbd} with $\cont{\pattern} = \plant{\pattern} = {\pattern}$ and with controller matrices satisfying $\eqref{eq:M_homo}$,  
	such that $\Plantclp = \LFTu({\Plant},\controrig)$ 
	is stable and has an 
	$\mathcal{L}_2$-gain 
	less than $\gamma$, if there exist $\mathcal{X}_i = \mathcal{X}_i^\trans > 0$, and 
	$\smallQ{R}_i= \smallQ{R}_i^\trans$, $\smallQ{Q}_i= \smallQ{Q}_i^\trans$ and $\smallQ{S}_i$, 
	such that 
	\eqref{eq:primalLMIs_small_hetero} and 
	\begin{equation}
	\label{eq:smallmultiplierspec_}
	\bma{@{}c@{}}{ \star }^T 
	\bma{@{}c@{\,\,\,\,\,\,} c@{}}{ 
		{\smallQ{Q}_i}  & {\smallQ{S}_i} \\
		{\smallQ{S}_i^\trans}  & {\smallQ{R}_i}}
	\bma{@{}c@{}}{ 
		\lambda I_{ n_{\interci}}\\
		I_{ n_{\interci}}  }
	> 0, ~\forall \lambda \in \mathrm{spec}\left({{\pattern}}\right). 
	\end{equation}
\end{myprop}

\begin{myrem}
	Note that the decomposed synthesis equations of Proposition~\ref{prop:smallLMIs_hetero} can also be used in the case of homogeneous systems. 
	Choosing identical multipliers $Q_{ik}$, $R_{ik}$ and $S_{ik}$, respectively, for all edges $(i,k)$, allows to decompose the nominal and multiplier conditions to identical small ones of the size of the individual subsystems. 
	However, as less structural knowledge about the interconnected system is exploited, in general more conservatism can be introduced by the controller design in Proposition~\ref{prop:smallLMIs_hetero} than by the one in Proposition~\ref{prop:smallLMIs_homo}. In particular, no information about the interconnection topology is captured in the system representation. 
	It can  be observed that the bound $\gamma$ of the resulting closed-loop under the synthesized controller from Proposition~\ref{prop:smallLMIs_hetero} is equal to the one from Proposition~\ref{prop:smallLMIs_homo} (and therefore no additional conservatism is introduced), only if the 
	graph is regular, and if the spectrum of the interconnection matrix is symmetric, i.e., if $\lambda_{\mathrm{min}}({\pattern}) = - \lambda_{\mathrm{max}}({\pattern})$. 
	If these conditions are not met, Proposition~\ref{prop:smallLMIs_homo} is less conservative than the synthesis based on Proposition~\ref{prop:smallLMIs_hetero} with identical multipliers for all edges. 
\end{myrem}

\subsection{Heterogeneous Groups of Subsystems with Different Homogeneous Interconnections}  
In between heterogeneous and homogeneous systems, we define a new class of systems with $\aalpha$ groups of homogeneous subsystems and with $\intercN$ different interconnection types, referred to as $\aalpha$-$\intercN$-heterogeneous systems. 
Technically, this classification includes heterogeneous systems as the most general case, with $\aalpha = N$ and $\intercN = \| \mathcal{E} \|$, and homogeneous systems as the most specific class, with $\aalpha = 1$ and $\intercN = 1$. The transformation of the system model and of the synthesis equations is most beneficial for small values of $\aalpha$ and $\intercN$. 

\begin{mydef}[$\aalpha$-$\intercN$-Heterogeneous systems]
	We define a system of $\aalpha$ groups of homogeneous subsystems with $\intercN$ different homogeneous interconnections, referred to as $\aalpha$-$\intercN$-heterogeneous systems, if its system matrices can be written as 
	\begin{equation}
	\label{eq:M_homohetero}
	M = \underbrace{ \sum_{i=1}^{\aalpha}\nolimits \kronecker{I_{\Theta_{i-1}+1:\Theta_i}}{ {M}_{ ii} } }_{\bd{M}} + 
	\underbrace{
		\sum_{i=1}^{\aalpha}\nolimits \sum_{j=1}^{\plant{\intercN}}\nolimits \left(  I_{\Theta_{i-1}+1:\Theta_i} \kronecker{ {\plant{\pattern}}_{j} }{{M}_{ij}}  \right) 
	}_{\obd{M}},
	\end{equation}
	where ${\plant{\pattern}}_{j}$ are different interconnection matrices, 
	and $I_{\Theta_{i-1}+1:\Theta_i}$ is an $N \times N$ matrix of all zeros except for the diagonal entries corresponding to the indices from $\Theta_{i-1}+1$ to $\Theta_i$ being ones. The index set variable $\Theta_i$ is defined as $\Theta_i = \sum_{l=1}^{i}\nolimits N_l$ with $\Theta_0 = 0$, where $N_l$ is the number of subsystems in the group $l$. 
	
	This means that within each of the $\aalpha$ groups, all subsystems have equal matrices $M_{ii}$ and can have $\plant{\intercN}$ different matrices, $M_{ij}$, interconnected through the interconnection matrices ${\plant{\pattern}}_{j}$. 
\end{mydef}

\begin{figure}
	\begin{tiny}
		\def\svgwidth{0.6\columnwidth}
		\centering \includegraphics[width=0.25\columnwidth]{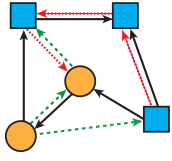} 
		\caption{\small Two groups of homogeneous subsystems ($\plant{\aalpha} = 2$) interconnected by three different interconnections ($\plant{\intercN} = 3$) symbolized by the different arrow types. 
			\label{fig:sysstruct}}
	\end{tiny}
\end{figure}
Figure~\ref{fig:sysstruct} shows an example with $\plant{\aalpha} = 2$ groups of homogeneous subsystems and $\plant{\intercN} = 3$ different interconnections. 
The matrices ${M}_{ij}$ correspond to those off-diagonal blocks of $M$ which represent the influence from all subsystems specified by the structure of ${\plant{\pattern}}_{j}$ to the subsystems $i$.

\begin{myprop}
If the controller $\controrig$ is also chosen to be composed of groups of homogeneous subcontrollers with different homogeneous interconnections, i.e., such that \eqref{eq:M_homohetero} holds with $M$ representing the controller matrices $\AK$, $\BK$, $\CK$ and $\DK$, then the interconnected closed loop system in  \eqref{eq:closed_loop_PPK_i} can be transformed to a representation where 
the interconnection matrix takes the form 
\begin{equation} \label{eq:P_homohetero}
\barbelow{\permute}
= 
\diagIndtwo{j=1}{\intercN}{\kronecker{{{\pattern}}_{j}}{{I_{n_{\interci_j}}}}}, 
\end{equation}
where $\intercN$ is the number of different interconnection matrices ${\plant{\pattern}}_{j}$ and ${\cont{\pattern}}_{j}$ in the closed-loop system. 
\end{myprop}
\begin{IEEEproof}
	We define the transformations of the interconnection channel and of the decentralized system part as 
	$\barbelow{{\permute}} = Z^\trans {\permute} Z$ and $\barbelow{\bd{\Plantclp}} = Z^{-1}  \bd{\Plantclp}  Z^{-\trans}$, respectively. 
	The transformation matrix $Z$ is defined as the $\vert \mathcal{E} \vert \times (\intercN \, N)$ concatenated matrix $Z = \concatIndtwo{j=1}{\intercN}{Z_j}$, where $Z_j$ are the $\vert \mathcal{E} \vert \times N$ matrices of all zeros except for ones in the entries corresponding to an interconnection defined in ${{\pattern}}_{j}$ between an edge and a system. With this transformation, the $\intercN$ interconnection topologies are captured in $\barbelow{{\permute}} = \mathrm{diag}(\kronecker{{{\pattern}}_{j}}{I_{n_{\interci}}})$.  
\end{IEEEproof}
This transformation is shown in Figure~\ref{fig:interconn_trafo} and is illustrated in Example~\ref{exp:trafo_homohetero}. A possible realization of the transformed system matrices of $\barbelow{\bd{\Plantclp}}$ are given in the Appendix in \eqref{eq:interconn_matrices_homohetero}.

\begin{myrem}
As in the homogeneous case, instead of stacking multiple identical interconnection channels for neighboring subsystems, they are summarized as one interconnection channel for each subsystem.  
In order to transform the representation to this form, the interconnection channel of each subsystem is augmented (by zero signals) such that each subsystem has interconnection signals belonging to all different interconnection matrices ${{\pattern}}_{j}$. 
Therefore, this formulation in general involves a larger dimension of the interconnection channel. It thus depends on the degree of homogeneity of the system, i.e., on  $\aalpha$ and $\intercN$, whether this formulation is beneficial in comparison to Proposition~\ref{prop:smallLMIs_hetero}. 
\end{myrem}

\begin{myrem}
	The modeling approach in \eqref{eq:M_homohetero} extends the special case of $\aalpha$-heterogeneous systems in \cite{Massioni2014}, where different groups of homogeneous subsystems are connected by identical interconnections. 
	The formulation given here captures the more general case of groups of homogeneous subsystems with different interconnections. 
	In this case, for example, the states can be interconnected through a different interconnection matrices than the performance inputs or outputs, or the control inputs. An example is given in \cite{Stuerz2018a} resulting from an augmented overlapping system representation as presented in \cite{Stuerz2017a} and analyzed in \cite{Stuerz2017b}. 
\end{myrem}

We will show in the following how the controller synthesis can be decoupled and simplified based on the system representation in \eqref{eq:M_homohetero}.

\subsection{Decomposed Controller Synthesis for $\aalpha$-$\intercN$-Heterogeneous Systems} 
\begin{myprop}
	\label{prop:smallLMIs_homohetero}
	\textit{(Decomposed Full-Block S-Procedure for $\aalpha$-$\intercN$-Heterogeneous Systems:)} 
	Let us consider the system $\Plant  = \LFTl(\bd{\Plant},\plant{\permute})$ given in \eqref{eq:interconn_openloop}, with system matrices as defined in \eqref{eq:M_homohetero},  
	Then there exists a controller $\controrig$ 
	as in \eqref{eq:interconn_control_i} 
	with controller matrices satisfying \eqref{eq:M_homohetero}, such that the interconnection matrix of the closed-loop system is given in \eqref{eq:P_homohetero}, 
	such that $\Plantclp = \LFTu(\Plant,\controrig)$ 
	is stable and has an 
	$\mathcal{L}_2$-gain  
	less than $\gamma$, if there exist ${\mathcal{X}}_i = {\mathcal{X}}_i^\trans > 0$, and 
	$\smallQ{Q}_j = \smallQ{Q}_j^\trans$, $\smallQ{R}_j= \smallQ{R}_j^\trans$ and $\smallQ{S}_j$, $\forall j = \{1,...,\intercN\}$, 
	with $\smallQ{Q}= \diagIndtwo{j=1}{\intercN}{\smallQ{Q}_j}$, $\smallQ{R}= \diagIndtwo{j=1}{\intercN}{\smallQ{R}_j}$ and $\smallQ{S}= \diagIndtwo{j=1}{\intercN}{\smallQ{S}_j}$ such that
	\begin{align}
	\label{eq:primalLMIs_small_mult_homohetero}
	\small \bma{@{}c@{}}{ \star }^T 
	\small \bma{@{}c@{\,\,\,\,\,\,} c@{}}{ 
		\kronecker{I_N}{\smallQ{Q}_{j}}  & \kronecker{I_N}{\smallQ{S}_{j}} \\
		\kronecker{I_N}{\smallQ{S}_{j}^\trans}  & \kronecker{I_N}{\smallQ{R}_{j}}} 
	\small \bma{@{}c@{}}{ 
		\kronecker{{{\pattern}}_{j}}{I_{n_{\interci_j}}} \\
		I_{N n_{\interci_j}}  } &>0, \notag\\
	\forall j \in \{1,...,\intercN\}, & \\
	\bma{@{}c@{}}{ 
		\star
	}^\trans 
	\small \bma{@{}c@{\,\,\,\,\,} c@{\,\,\,\,\,} : c@{\,\,\,\,\,}  c@{\,\,\,\,\,} : c@{}  c@{}}{ 
		0 & {\mathcal{X}}_i  & 0 & 0 & 0 & 0 \\
		{\mathcal{X}}_i  & 0 & 0 & 0 & 0 & 0 \\
		\hdashline
		0 & 0 &  - \gamma I & 0 & 0 & 0 \\
		0 & 0 & 0 & \frac{1}{\gamma} I & 0 & 0 \\
		\hdashline
		0 & 0 & 0 & 0 & \smallQ{Q}  & \smallQ{S}  \\
		0 & 0 & 0 & 0 &  \smallQ{S}^\trans  & \smallQ{R} 
	} 
	\small \bma{@{}c@{\,\,\,\,\,\,} c@{\,\,\,\,\,\,}  c@{}}{ 
		I & 0 & 0 \\
		\mathcal{A}_i & \mathcal{B}_{1,i} & \mathcal{B}_{2,i} \\
		\hdashline
		0 & I & 0 \\
		\mathcal{C}_{1,i} & \mathcal{D}_{11,i} & \mathcal{D}_{12,i} \\
		\hdashline
		0 & 0 & I \\
		\mathcal{C}_{2,i} & \mathcal{D}_{21,i} & \mathcal{D}_{22,i}
	} 
	&< 0 , \notag\\
	\forall i \in \{1,...,\aalpha\}. &
	\label{eq:primalLMIs_small_homohetero}
	\end{align}
\end{myprop}
\begin{IEEEproof}
	The proof follows along the same lines as the one of Proposition~\ref{prop:smallLMIs_hetero} with 
	the structured Lyapunov matrix $\mathcal{X} = \diagIndtwo{i=1}{\aalpha}{\kronecker{I_{N_i}}{\mathcal{X}}_i}$ and the structured multipliers 
	$Q = \diagIndtwo{j=1}{\intercN}{\kronecker{I_N}{\smallQ{Q}}_{j}}$, $R = \diagIndtwo{j=1}{\intercN}{\kronecker{I_N}{\smallQ{R}}_{j}}$ and $S = \diagIndtwo{j=1}{\intercN}{\kronecker{I_N}{\smallQ{S}}_{j}}$. 
	Herein, $N_i$ is the number of subsystems within the group $i$. 
		Note that if the system $\Plant$ is the transformed system $\bar{\Plant}$ in \eqref{eq:overallsys_orig}, Corollary \ref{cor:decSproc_GbarG} applies such that Proposition~\ref{prop:smallLMIs_homohetero} also holds for $\bar{\Plantclp}$.  
\end{IEEEproof}

\begin{mycor} \label{cor:lambda_j}
	Furthermore, by applying Lemma~1 from \cite{Hoffmann2016}, we can state the following. 
	If the interconnection matrices $P_j$ are normal, they can always be transformed into diagonal matrices with their eigenvalues on the diagonal. 	
	Introducing the additional constraint $\tilde{Q}_j > 0$ guarantees concavity of the multiplier condition in $\lambda$, which lets us further decompose the multiplier condition in \eqref{eq:primalLMIs_small_mult_homohetero} into the following conditions. 
	\begin{equation}
	\label{eq:smallmultiplierspec}
	\bma{@{}c@{}}{ \star }^T 
	\bma{@{}c@{\,\,\,\,\,\,} c@{}}{ 
		{\smallQ{Q}_j}  & {\smallQ{S}_j} \\
		{\smallQ{S}_j^\trans}  & {\smallQ{R}_j}}
	\bma{@{}c@{}}{ 
		\lambda I_{ n_{\interci_j}}\\
		I_{ n_{\interci_j}}  }
	> 0, ~\forall \lambda \in \mathrm{spec}\left({{{\pattern}}}_{j}\right), \\
	\forall j \in \{1,...,\intercN\}. 
	\end{equation}
\end{mycor}

\begin{myrem} 
	The same observation as in Remark~\ref{rem:SFOF} holds here. 
	As for the synthesis equations in Proposition~\ref{prop:smallLMIs_hetero}, also the synthesis conditions in Proposition~\ref{prop:smallLMIs_homohetero} are convex in the case of state feedback. In the case of dynamic output feedback, a variable transformation leads to bilinear matrix inequalities which can be handled by iteratively solving two LMIs.  
\end{myrem}

Thus, in the case of $\aalpha$ groups of homogeneous subsystems, there are $\aalpha$ small nominal conditions to be solved. 
Furthermore, for each of the $\intercN$ interconnections, two multiplier conditions (for the smallest and largest eigenvalues of the ${{\pattern}}_{j}$), need to be solved.

\begin{myrem}
	In \cite{Massioni2014}, a decomposition of the controller synthesis equations for $\aalpha$-heterogeneous systems based on a singular-value decomposition is proposed, which potentially introduces more conservatism as congruence transformation proposed in Proposition~\ref{prop:smallLMIs_homohetero}.  
\end{myrem}

\section{Numerical Example}
\label{sec:NumericalExample}

\subsection{Illustration of the Theory}

 The following examples will illustrate the definition and construction of the interconnection matrices. 
 
\begin{myexp}[Definition of interconnections $\interco, \interci$ and interconnection matrix $\permute$]
	\label{exp:trafo}
	We consider the following closed-loop system composed of four subsystems with the interconnection matrix 
	${{\pattern} = \scriptsize \bma{@{} c@{\,\,\,\,\,\,} c@{\,\,\,\,\,\,} c@{\,\,\,\,\,\,} c@{}}{0 & \pattern_{12} & 0 & 0 \\ \pattern_{21} & 0 & \pattern_{23} & \pattern_{24} \\ 0 & \pattern_{32} & 0 & 0 \\ 0 & \pattern_{42} & 0 & 0}}$ indicating the interconnection topology of the subsystems.  
	The interconnection channel, as defined in \eqref{eq:interc_signals_def}, for the closed-loop system is 
	$$\underbrace{\scriptsize \bma{@{} c@{}}{\interci_{12}\\ \interci_{21}\\ \interci_{23}\\ \interci_{24}\\ \interci_{32}\\ \interci_{42}}}_{\interci} = 
	\underbrace{\scriptsize \bma{@{} c@{\,\,\,\,\,\,} : c@{\,\,\,\,\,\,} c@{\,\,\,\,\,\,} c@{\,\,\,\,\,\,} : c@{\,\,\,\,\,\,} : c@{}}{
			& \permute_{12} & & & & \\
			\hdashline
			\permute_{21} & & & & & \\
			& & & & \permute_{23}& \\	
			& & & & & \permute_{24} \\
			\hdashline	
			& & \permute_{32}& & & \\
			\hdashline	
			& & & \permute_{42}& & 
	}}_{\permute}
	\underbrace{\scriptsize \bma{@{} c@{}}{\interco_{12}\\ \interco_{21}\\ \interco_{23}\\ \interco_{24}\\ \interco_{32}\\ \interco_{42}}}_{\interco}.$$ 
For example, subsystem 1 has an interconnection channel of $n_{\interci_1} = n_{\interci_{12}}$ to $n_{\interco_1} = n_{\interco_{12}}$ and 
	subsystem 2 has an interconnection channel of $n_{\interci_2} = n_{\interci_{21}} + n_{\interci_{23}} + n_{\interci_{24}}$ to $n_{\interco_2} = n_{\interco_{21}} + n_{\interco_{23}} + n_{\interco_{24}}$. 
\end{myexp}

\begin{myexp}[Transformation to block-diagonal $\permute$] \label{exp:trafo_blkdiag}
	For the system in Example~\ref{exp:trafo}, it is easy to see that the transformation $T = \scriptsize \bma{@{} c@{\,\,\,\,\,\,} c@{\,\,\,\,\,\,} c@{\,\,\,\,\,\,} c@{\,\,\,\,\,\,} c@{\,\,\,\,\,\,} c@{}}{
	   I&  & & & & \\
		& I& & & & \\
		& & I& & & \\	
		& & & & I& \\	
		& & & I& & \\	
		& & & & & I
	},$ 
applied to the system as in \eqref{eq:trafoT} gives 
	$$\underbrace{\scriptsize \bma{@{} c@{}}{\interci_{12} \\ \interci_{21} \\ \interci_{23} \\ \interci_{32} \\ \interci_{24} \\ \interci_{42}}}_{\bar{\interci}} = 
	\underbrace{\scriptsize \bma{@{} c@{\,\,\,\,\,\,} c@{\,\,\,\,\,\,} : c@{\,\,\,\,\,\,} c@{\,\,\,\,\,\,} : c@{\,\,\,\,\,\,} c@{}}{
			& \permute_{12} & & & & \\
			\permute_{21} & & & & & \\
				\hdashline
			& & & \permute_{23}& & \\	
			& & \permute_{32}& & & \\
				\hdashline	
			& & & & & \permute_{24}\\	
			& & & & \permute_{42}& 
	}}_{\bar{\permute}}
	\underbrace{ \scriptsize \bma{@{} c@{}}{\interco_{12} \\ \interco_{21} \\ \interco_{23} \\ \interco_{32} \\ \interco_{24} \\ \interco_{42}}}_{\bar{\interco}}.$$
\end{myexp}

\begin{myexp}[Transformation to simplified representation for homogeneous systems]
\label{exp:trafo_homo} 
If the system in Example~\ref{exp:trafo} is a homogeneous system, then it holds that all the interconnection channels over the different edges of a subsystem are identical, as all off-block-diagonal matrices $M_{ik}$ are identical, and therefore we can apply the transformation 
$$Z = 
\scriptsize \bma{@{} c@{\,\,\,\,\,\,}   c@{\,\,\,\,\,\,} c@{\,\,\,\,\,\,}  c@{}}
{I & 0 & 0 & 0 \\ 
	0 & I & 0 & 0  \\ 
	0 & I & 0 & 0  \\ 
	0 & I & 0 & 0  \\ 
	0 & 0 & I & 0 \\ 
	0 & 0 & 0 & I},$$ 
which leads to the interconnection matrix 
\begin{equation}\label{eq:permutehomo}
\begin{aligned}
\barbelow{\permute} &=  Z^{\trans} \permute Z = 
\scriptsize \bma{@{} c@{\,\,\,\,\,\,}   c@{\,\,\,\,\,\,} c@{\,\,\,\,\,\,}  c@{}}
{0 & \permute_{12} & 0 & 0 \\ 
	\permute_{21} & 0 & \permute_{23} & \permute_{24}  \\ 
	0 & \permute_{32} & 0 & 0 \\ 
	0 & \permute_{42} & 0 & 0} \\
&= \kronecker{{\pattern}}{I_{n_\interci}},  
\end{aligned} 
\end{equation}
which captures the information about the interconnection topology of the subsystems and summarizes all interconnection signals from different subsystems to the same subsystem in one channel. 
The last equality in \eqref{eq:permutehomo} assumes that the individual interconnection matrices $\permute_{ik}$ can be captured in the form involving the interconnection matrix $\kronecker{\pattern_{ik}}{I_{n_{\interci}}}$. 
The decentralized part of the system, $\bd{\Plantclp}$ is transformed according to the transformation in \eqref{eq:Ztransform}, such that the closed-loop system is not changed. 
\end{myexp}

\begin{myexp}[Transformation to simplified representation for $\aalpha$-$\intercN$-heterogeneous systems]
	\label{exp:trafo_homohetero}

Let us assume that subsystems 1 and 2 have equal diagonal (closed-loop) matrices, i.e., $M_{11} = M_{22} =: M_{1}$ and therefore form a homogeneous group and subsystems 3 and 4 form another one, i.e., $M_{33} = M_{44} =: M_{2}$, and thus $\aalpha = 2$. Furthermore, we assume that $M_{12}$ and $M_{21}$ are equal and form one group of homogeneous interconnections, and $M_{23}$, $M_{32}$, $M_{24}$ and $M_{42}$ are equal and form another one, and thus $\intercN = 2$. 
	The system can be modeled by the interconnection matrices  
	$${{{\pattern}}_{1} = 
	\scriptsize \bma{@{} c@{\,\,\,\,\,\,} c@{\,\,\,\,\,\,} c@{\,\,\,\,\,\,} c@{}}{0 & \pattern_{12} & 0 & 0 \\ \pattern_{21} & 0 & 0 & 0 \\ 0 & 0 & 0 & 0 \\ 0 & 0 & 0 & 0},  \quad 
	{{\pattern}}_{2} = 
	\scriptsize \bma{@{} c@{\,\,\,\,\,\,} c@{\,\,\,\,\,\,} c@{\,\,\,\,\,\,} c@{}}{0 & 0 & 0 & 0 \\ 0 & 0 & \pattern_{23} & \pattern_{24} \\ 0 & \pattern_{32} & 0 & 0 \\ 0 & \pattern_{42} & 0 & 0}.}$$ 

We can apply the following transformation 
$$Z =  
\scriptsize \bma{@{} c@{\,\,\,\,\,\,}   c@{\,\,\,\,\,\,} c@{\,\,\,\,\,\,} c@{\,\,\,\,\,\,} : c@{\,\,\,\,\,\,} c@{\,\,\,\,\,\,} c@{\,\,\,\,\,\,}  c@{}}
{I &  &  &  & 0 &  &  &  \\ 
  & I &  &  &  & 0 &  &  \\ 
  & 0 &  &  &  & I &  &  \\ 
  & 0 &  &  &  & I &  &  \\ 
  &  & 0 &  &  &  & I &  \\ 
  &  &  & 0 &  &  &  & I},$$ 
which leads to the interconnection matrix 
\begin{equation}
\begin{aligned}
\barbelow{\permute} &=  Z^{\trans} \permute Z = 
\scriptsize \bma{@{} c@{\,\,} c@{\,\,\,\,}  c@{\,\,\,\,\,\,\,\,} c@{\,\,\,\,\,\,\,\,}  : c@{\,\,\,\,\,\,} c@{\,\,}  c@{\,\,}  c@{}}
{0 & \permute_{12} & 0 &  0 & 0 & 0 & 0 & 0  \\ 
	\permute_{21} & 0 & 0 & 0 & 0 & 0 & 0 & 0  \\ 
	0 &	0 & 0 & 0 & 0 & 0 & 0 & 0  \\
	0 & 0 & 0 & 0 & 0 & 0 & 0 & 0  \\ 
	\hdashline
	0 & 0 & 0 & 0 & 0 & 0 & 0 & 0  \\ 
	0 & 0 & 0 & 0 & 0 & 0 & \permute_{23} & \permute_{24}  \\ 
	0 & 0 & 0 & 0 & 0 & \permute_{32} & 0 & 0  \\ 
	0 & 0 & 0 & 0 & 0 & \permute_{42} & 0 & 0  } \\
&=
\mathrm{diag}( \kronecker{{{\pattern}}_{1}}{I_{n_{\interci_1}}}, \kronecker{{{\pattern}}_{2}}{I_{n_{\interci_2}}}) 
=
\diagIndtwo{j=1}{\intercN}{\kronecker{{{\pattern}}_{j}}{I_{n_{\interci_j}}}}, 
\end{aligned}
\end{equation}
which is expressed by the interconnection matrices ${{\pattern}}_{j}$. 
The last equality again uses the assumption that the individual interconnection matrices $\permute_{ik}$ can be expressed in the form $\kronecker{\pattern_{ik}}{I_{n_{\interci}}}$.  
The decentralized part of the system, $\bd{\Plantclp}$ is transformed according to the transformation in \eqref{eq:Ztransform}, such that the closed-loop system is not changed. 
Note that this modeling is not unique. The number $\intercN$ is at least the maximum of the numbers of different interconnections over the subsystems. 
\end{myexp}

\subsection{Example Systems}
In the following, we consider randomly generated example systems, based on coupled mass-spring-damper subsystems. Each subsystem has a mass $m_i \in \mathcal{U}(5,10)$, spring and damping coefficients $k_i \in \mathcal{U}(0.8,1.2)$ and $d_i \in \mathcal{U}(0.8,1.2)$, respectively, where $\mathcal{U}(a,b)$ denotes the uniform distribution with support on the interval $[a,b]$. The interconnections between the subsystems are described by the spring and damping coupling coefficients, $k_{ik} \in \mathcal{U}(0.2,0.4)$ and $d_{ik} \in \mathcal{U}(0.2,0.4)$. 
The system matrices are given by 
\begin{equation} \label{eq:randomsys}
\begin{aligned}
A_{ii} &= \bma{@{}c@{\,\,\,\,}c@{}}{0 & 1 \\ - \frac{\sum_{k \in \mathcal{N}_i}{k_{ik}}}{m_i} & - \frac{\sum_{k \in \mathcal{N}_i}{d_{ik}}}{m_i}}, \quad A_{ik} = \bma{@{}c@{\,\,\,\,}c@{}}{0 & 0 \\ \frac{{k_{ik}}}{m_k} & \frac{{d_{ik}}}{m_k}}, \\ 
B_{u,i} &= \bma{@{}c@{}}{0 \\ b_{u,i}}, ~~ D_{\perfo u,i} = \bma{@{}c@{}}{0_{n_{xi} \times n_{ui}} \\ d_{\perfo u,i}}, ~~ \text{with} ~ b_{u,i}, ~ d_{\perfo u,i} \in \mathcal{U}(1,1.3), \\
C_{y,i} &= I, \quad  C_{\perfo,i} = \bma{@{}c@{}}{I \\ 0_{n_{\perfi i}-n_{xi},n_{xi}}}, \\
B_{\perfi,i} &= \bma{@{}c@{}}{0 \\ b_{\perfi,i}}, ~~ \text{with} ~ b_{\perfi,i}, \in \mathcal{U}(1,1.2), 
\end{aligned}
\end{equation}
and the remaining system matrices are zero.

\subsection{Convergence Results of ADMM}

\begin{figure}
	\begin{tiny}
		\def\svgwidth{0.7\columnwidth}
		\centering
		\input{figures/directedgraph.tex}
		\includegraphics[width=0.7\columnwidth]{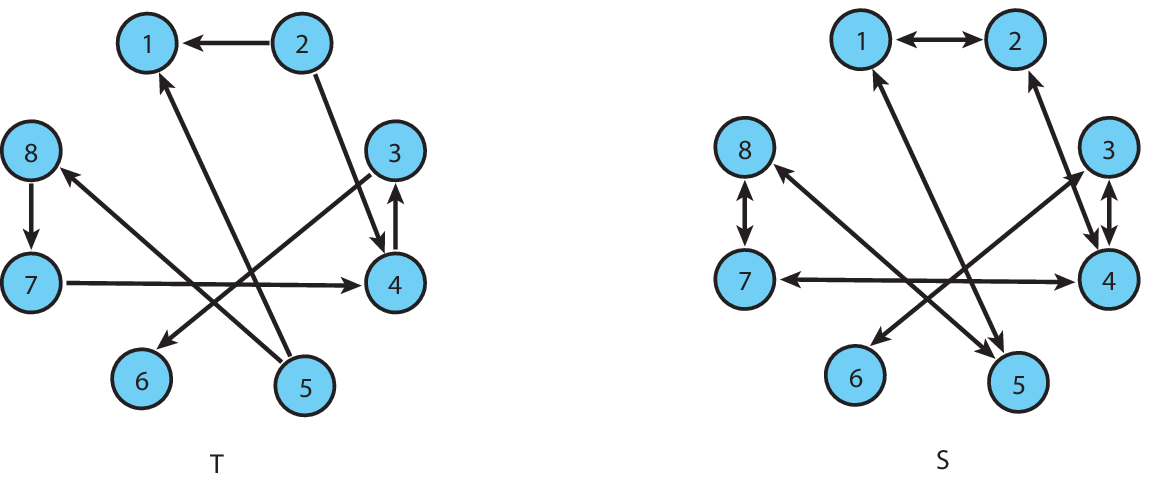} 
		\caption{\small Interaction graphs of the system and the controller $\plant{\mathcal{E}} = \cont{\mathcal{E}}$ and communication graph for distributed control design $\mathcal{E} = \plant{\mathcal{E}} \cup \plantmirror{\mathcal{E}} \cup \cont{\mathcal{E}} \cup \contmirror{\mathcal{E}}$. 
			\label{fig:directedgraph}}
	\end{tiny}
\end{figure}

\begin{figure*}
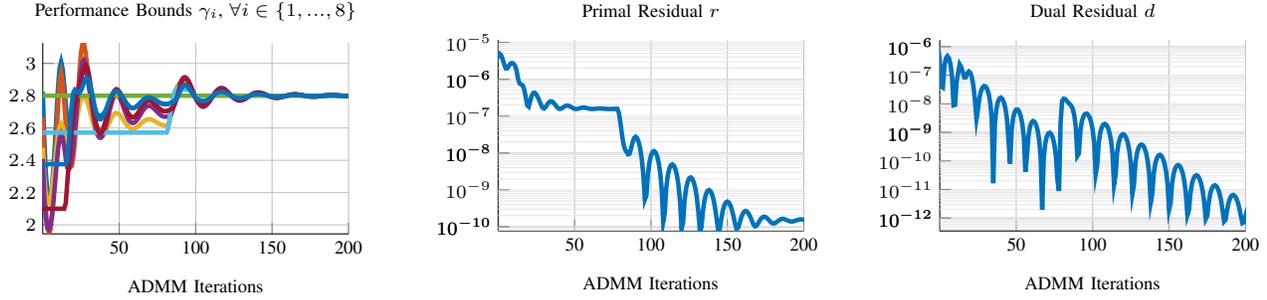

	\begin{scriptsize}
		\captionsetup[subfigure]{aboveskip=-1pt,belowskip=-1pt}
		\begin{subfigure}{0.3\textwidth}
			\begin{center}
				\input{figures/TAC_N8dist2_TAC_Rev_gamma_tikz_.tex}
			\end{center}
		\end{subfigure}
		~~~
		\begin{subfigure}{0.3\textwidth}
			\begin{center}
				\input{figures/TAC_N8dist2_TAC_Rev_residualsr_tikz_.tex}
			\end{center}
		\end{subfigure}
		~~~
		\begin{subfigure}{0.3\textwidth}
			\begin{center}
				\input{figures/TAC_N8dist2_TAC_Rev_residualsd_tikz_.tex}
			\end{center}
		\end{subfigure}
		\caption{\small Convergence results for the ADMM scheme in Algorithm~\ref{alg:ADMM} for an example  system of 8 interconnected subsystems. \label{fig:ADMMconvergence}}
	\end{scriptsize} 
\end{figure*}

We present the convergence of the ADMM scheme in Algorithm~\ref{alg:ADMM} for an example system containing $N=8$ interconnected subsystems with matrices randomly chosen as given in \eqref{eq:randomsys}. The interconnection topology is $\plant{\mathcal{E}} = \{$ $(1,5)$, $(2,1)$, $(3,4)$, $(4,2)$,  $(4,7)$, $(5,6)$, $(6,3)$, $(7,8)$, $(8,5)$ $\}$. 
We consider the convex synthesis of interconnected static state feedback controllers, which has the same interconnection structure as the system, i.e., $\cont{\mathcal{E}}=\plant{\mathcal{E}}$. 
For the distributed controller synthesis, the communication topology is given by $\mathcal{E} = \plant{\mathcal{E}} \cup \plantmirror{\mathcal{E}} \cup \cont{\mathcal{E}} \cup \contmirror{\mathcal{E}}$, as shown in Figure~\ref{fig:directedgraph}. 
Figure~\ref{fig:ADMMconvergence} shows the convergence of the bound $\gamma$ on the $\mathcal{H}_\infty$-norm of the example system and the convergence of the primal and dual residuals given in \eqref{eq:residuals}. 

This convergence behavior is representative for the class of interconnected systems. The convergence time and the oscillatory behavior however depend on the number of subsystems and on the interconnection topology.

\subsection{Scalability of the Centralized Synthesis and the Decomposed Synthesis for Homogeneous and Heterogeneous Systems}

In the following, we compare the computational scalability of the centralized synthesis with full block multipliers for heterogeneous systems in Theorem~\ref{the:sproc} 
with the decomposed synthesis for heterogeneous systems in Proposition~\ref{prop:smallLMIs_hetero},  and 
for the special cases of homogeneous systems and groups of homogeneous subsystems in Propositions~\ref{prop:smallLMIs_homo} and \ref{prop:smallLMIs_homohetero}, respectively, for a growing number of subsystems, $N$, and groups $\aalpha$, respectively. 

For comparability, these problems are all solved in a centralized way, i.e., on one computer in one thread. Their scalability for a growing number of subsystems, $N$, or groups of homogeneous subsystems, $\aalpha$, and for a growing number of interconnections $\vert\mathcal{E}\vert$, is investigated in terms of LMI size, optimization variables  solver times. 

A direct comparison of the computational scalability of the centralized decomposed synthesis methods with the distributed synthesis method in Algorithm~\ref{alg:ADMM} in terms of solver time is not possible, because the convergence time of the distributed synthesis heavily depends on the interconnection topology of the system. 
Therefore, we compare the scalability of the centralized decomposed synthesis methods to the computational effort for one subsystem in one iteration of the distributed synthesis.

We consider the worst case interconnection topology w.r.t.\ computational scalability, i.e., the case where all subsystems are interconnected with all other subsystems. The system matrices are chosen as in \eqref{eq:randomsys}.

\begin{table}
	\centering 
	\begin{small}
		\begin{tabular}{@{}l@{} || @{}c@{} | @{}c@{} || @{}c@{} | @{}c@{} }
			& \multicolumn{2}{c||}{ \parbox{2.38cm}{~\\[-0.05cm] Nominal condition\\[-0.05cm] ~}} & \multicolumn{2}{c}{\parbox{2.6cm}{~\\[-0.05cm] Multiplier condition\\[-0.05cm] ~}} \\
			\cline{2-5} 
			\centering \parbox{1.9cm}{ Synthesis \\[-0.05cm] ~} &  \, Number \, & Size  & \, Number \, & Size  \\
			\hline 
			\parbox{1.9cm}{ ~\\[-0.05cm] Centralized \\[-0.05cm]~} &  1 &  \, $N \bar{X}_{\mathrm{n}} \! \times \! N \bar{X}_{\mathrm{n}}$ \, &  1 & \, $\vert \mathcal{E} \vert X_{\mathrm{m}} \! \times \! \vert \mathcal{E} \vert X_{\mathrm{m}}$  \\
			\parbox{1.9cm}{ Decomposed \\[-0.1cm] homogeneous} &   1 &  $X_{\mathrm{n}} \! \times \! X_{\mathrm{n}}$ &  $2$ & $ X_{\mathrm{m}} \! \times \! X_{\mathrm{m}}$ \\[0.2cm]
			%
			\parbox{1.9cm}{ Decomposed \\[-0.1cm] heterogeneous\\[-0.1cm] groups } &   $\aalpha$ & $X_{\mathrm{n}} \! \times \! X_{\mathrm{n}}$  &   $2\beta$ & $X_{\mathrm{m}} \! \times \! X_{\mathrm{m}}$ \\[0.3cm] 
			%
			\parbox{1.9cm}{ Decomposed \\[-0.1cm] heterogeneous} &   $N$ & $\bar{X}_{\mathrm{n}} \! \times \! \bar{X}_{\mathrm{n}}$  &   \! $\vert \mathcal{E} \vert$ & $X_{\mathrm{m}} \! \times \! X_{\mathrm{m}}$  \\[0.3cm] 
			%
			\parbox{2.0cm}{ Distributed \\[-0.1cm] heterogeneous$^{\ast}$ } &    \centering $1$ & $\bar{X}_{\mathrm{n}} \! \times \! \bar{X}_{\mathrm{n}}$ &   \centering \! $\vert \mathcal{N}_i \vert$ & $X_{\mathrm{m}} \! \times \! X_{\mathrm{m}}$ \\[0.1cm] 
		\end{tabular}
	\end{small}
	\caption{\small 
		Numbers and dimensions of synthesis conditions for the centralized (Theorem~\ref{the:sproc}) and decomposed synthesis for homogeneous subsystems (Proposition~\ref{prop:smallLMIs_homo}), $\aalpha$-$\intercN$-heterogeneous systems (Proposition~\ref{prop:smallLMIs_homohetero}), and heterogeneous subsystems (Proposition~\ref{prop:smallLMIs_hetero}), with the dimensions of the (mean) nominal, $X_\mathrm{n}$ ($\bar{X}_\mathrm{n}$), and the (mean) multiplier conditions, $X_\mathrm{m}$ ($\bar{X}_\mathrm{m}$), for the single subsystems. Extended from \cite{Stuerz2018a}. \quad
		$^{\ast}$ Numbers and sizes of LMIs are given per subsystem $i$ and per iteration of Algorithm~\ref{alg:ADMM}. \qquad \quad
		\label{tab:scale}	
	}
\end{table}
Table~\ref{tab:scale} shows the number and dimensions of LMIs to be solved for the centralized and the decomposed controller syntheses, and for the distributed synthesis per subsystem and ADMM iteration. For simplicity, we assume that the dimensions of the single subsystems are equal although they can be heterogeneous. We denote by $X_\mathrm{n}$ the dimension of one of the small decomposed nominal conditions of the dimension of one subsystem, and by $X_\mathrm{m}$ the dimension of one small decomposed multiplier condition of the dimension of one subsystem.  Note that this is a simplification, as in general, the dimensions of the nominal and the multiplier conditions for the different formulations in Propositions~\ref{prop:smallLMIs_hetero}, \ref{prop:smallLMIs_homo}, and \ref{prop:smallLMIs_homohetero}, are not equal, but also depend on the number of neighboring subsystems (in Propositions~\ref{prop:smallLMIs_hetero} and \ref{prop:smallLMIs_homohetero}). This is indicated by $\bar{X}_\mathrm{n}$ which thus indicates the mean value of the size of the conditions. 

While the centralized synthesis scales polynomially with both the number of subsystems $N$ and the number of edges $\vert \mathcal{E} \vert$, 
the decomposed approach for heterogeneous systems scales linearly in both the number of subsystems $N$ and the number of edges $\vert \mathcal{E} \vert$. 
In the special case of groups of homogeneous subsystems this scaling is linear in $\aalpha$ and $\intercN$, respectively. The factor 2 applies to the normal case accounting for the smallest and largest eigenvalues of $\clp{\pattern}$ or each ${\clp{\pattern}}_{\!\!j}$ as in Corollary~\ref{cor:lambda_j}. 
For homogeneous systems, the computational effort for the synthesis is constant, i.e., it does not depend on $N$ and $\vert\mathcal{E}\vert$.

For each subsystem in each iteration of the ADMM scheme in Algorithm~\ref{alg:ADMM}, the LMI size, the number of LMIs, the number of optimization variables, and the amount of communication, all scale linearly with the number of neighboring subsystems to which the respective subsystem is interconnected. All of these variables are independent of the total number of subsystems $N$. This result can also be seen in Table~\ref{tab:scale}. 

The number of optimization variables and the solver times, averaged over 10 computations, are shown in Figure~\ref{fig:scalestudy} on a logarithmic scale. 
Note that for the $\aalpha$-$\intercN$-heterogeneous system, we assume $\intercN=1$ and the scaling is shown over the number of groups $\aalpha$. Therefore, the heterogeneous system involves more optimization variables, since it does not only scale with the number of subsystems, but also with the number of neighboring subsystems. However, this is compensated by less coupling because of the more structured multipliers, which is why the solver times for both systems (in terms of one ADMM iteration per subsystem for the latter) are very similar. Also note that for the centralized synthesis of Theorem~\ref{the:sproc}, we chose the multipliers to be block-diagonal for the subsystems. Even with this simplification, the solver times rapidly become prohibitive.

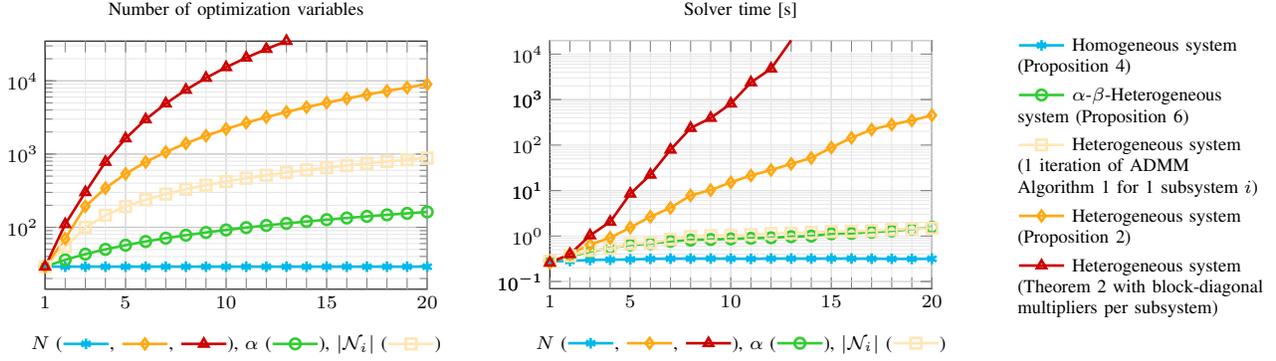
\begin{figure*}
	\begin{scriptsize}
		\begin{centering}
		\captionsetup[subfigure]{aboveskip=-1pt,belowskip=-1pt}
		\begin{subfigure}{0.36\textwidth}
			\begin{center}
%
%
\definecolor{mycolor1}{rgb}{0.00000,0.44700,0.74100}%
\begin{tikzpicture}

\begin{axis}[%
width=2.0in,
height=1.3in,
at={(0.409449in,0.418704in)},
scale only axis,
xmin=1,
xmax=20,
%
xlabel={$N$ (\ref{homo_Numvar}, \ref{heteroedgedecomp_Numvar}, \ref{heteroedge_centralized_blkdiagpersys_Numvar}), $\aalpha$ (\ref{homohetero_Numvar}), $\vert \mathcal{N}_i \vert$ (\ref{ADMM1iter_Numvar})},
ymin=0,
ymax=34881,
ymode=log,
log basis y={10},
xtick = {1,5,10,15,20},
grid=both,
xminorgrids,
yminorgrids,
grid style={line width=.1pt, draw=gray!20},
major grid style={line width=.2pt,draw=gray!50},
minor grid style={line width=.1pt,draw=gray!20},
minor x tick num=1,
extra x ticks={1,...,20},
extra x tick labels={},
extra tick style={
	tick style={line width=0.2pt},
	minor grid style={draw=gray!20},
	grid=minor,
},
minor y tick num=2,
title={Number of optimization variables},
legend style={legend cell align=left,align=left,draw=white!15!black}
]


%
%
%
\addplot [color=ProcessBlue,solid,line width=1.0pt,mark=asterisk,mark options={solid},forget plot]
table[row sep=crcr]{%
	1	29\\
	2	29\\
	3	29\\
	4	29\\
	5	29\\
	6	29\\
	7	29\\
	8	29\\
	9	29\\
	10	29\\
	11	29\\
	12	29\\
	13	29\\
	14	29\\
	15	29\\
	16	29\\
	17	29\\
	18	29\\
	19	29\\
	20	29\\
};\label{homo_Numvar}


\addplot [color=LimeGreen,solid,line width=1.0pt,mark=o,mark options={solid},draw=LimeGreen,forget plot]
table[row sep=crcr]{%
	1	29\\
	2	36\\
	3	43\\
	4	50\\
	5	57\\
	6	64\\
	7	71\\
	8	78\\
	9	85\\
	10	92\\
	11	99\\
	12	106\\
	13	113\\
	14	120\\
	15	127\\
	16	134\\
	17	141\\
	18	148\\
	19	155\\
	20	162\\
};\label{homohetero_Numvar}


\addplot [color=Peach,solid,line width=1.0pt,mark=square,forget plot]
table[row sep=crcr]{%
	1	29\\
	2	54\\
	3	100\\
	4	146\\
	5	192\\
	6	238\\
	7	284\\
	8	330\\
	9	376\\
	10	422\\
	11	468\\
	12	514\\
	13	560\\
	14	606\\
	15	652\\
	16	698\\
	17	744\\
	18	790\\
	19	836\\
	20	882\\
};\label{ADMM1iter_Numvar}


\addplot [color=darktangerine,solid,line width=1.0pt,mark=diamond,mark options={solid},forget plot]
table[row sep=crcr]{%
	1	29\\
	2	70\\
	3	193\\
	4	342\\
	5	537\\
	6	778\\
	7	1065\\
	8	1398\\
	9	1777\\
	10	2202\\
	11	2673\\
	12	3190\\
	13	3753\\
	14	4362\\
	15	5017\\
	16	5718\\
	17	6465\\
	18	7258\\
	19	8097\\
	20	8982\\
};\label{heteroedgedecomp_Numvar}

\addplot [color=Red,solid,line width=1.0pt,mark=triangle,mark options={solid},forget plot]
table[row sep=crcr]{%
1	29\\
2	110\\	
3	301\\
4	780\\
5	1633\\
6	2968\\
7	4893\\
8	7516\\
9	10945\\
10	15288\\
11	20653\\
12	27148\\
13	34881\\
};\label{heteroedge_centralized_blkdiagpersys_Numvar}



\end{axis}

\end{tikzpicture}%
			\end{center} 
		\end{subfigure} 
		\begin{subfigure}{0.36\textwidth}
			\begin{center}
%
%
\definecolor{mycolor1}{rgb}{0.00000,0.44700,0.74100}%
\begin{tikzpicture}
\begin{axis}[%
width=2.0in,
height=1.3in,
at={(0.409449in,0.418704in)},
scale only axis,
xmin=1,
xmax=20,
xlabel={$N$ (\ref{homo_Numvar}, \ref{heteroedgedecomp_Numvar}, \ref{heteroedge_centralized_blkdiagpersys_Numvar}), $\aalpha$ (\ref{homohetero_Numvar}), $\vert \mathcal{N}_i \vert$ (\ref{ADMM1iter_Numvar})},
ymin=0.07,
ymax=20037,
ymode=log,
log basis y={10},
xtick={1,5,10,15,20},
grid=both,
xminorgrids,
yminorgrids,
grid style={line width=.1pt, draw=gray!20},
major grid style={line width=.2pt,draw=gray!50},
minor grid style={line width=.1pt,draw=gray!20},
minor x tick num=1,
extra x ticks={1,...,20},
extra y ticks={1e-1, 1e0, 1e1, 1e2, 1e3, 1e4},
extra x tick labels={},
extra tick style={
	tick style={line width=0.2pt},
	minor grid style={draw=gray!20},
	grid=minor,
},
minor y tick num=2,
title={Solver time [s]},
legend style={legend cell align=left,align=left,draw=white!15!black}
]

\addplot [color=ProcessBlue,solid,line width=1.0pt,mark=asterisk,mark options={solid},forget plot]
table[row sep=crcr]{%
	1	0.27826334\\
	2	0.28203627\\
	3	0.29844515\\
	4	0.30193149\\
	5	0.30600387\\
	6	0.31524333\\
	7	0.31845362\\
	8	0.31904474\\
	9	0.31949693\\
	10	0.31851094\\
	11	0.31546683\\
	12	0.32028145\\
	13	0.3197704\\
	14	0.31892565\\
	15	0.32068046\\
	16	0.31927104\\
	17	0.31788381\\
	18	0.31516697\\
	19	0.31405983\\
	20	0.316541009999999\\
};\label{homo_solvetime}

\addplot [color=LimeGreen,solid,line width=1.0pt,mark=o,mark options={solid},forget plot]
table[row sep=crcr]{%
	1	0.27826334\\
	2	0.356883\\
	3	0.453755\\
	4	0.54935\\
	5	0.638081999999999\\
	6	0.681654999999999\\
	7	0.767393999999999\\
	8	0.817304\\
	9	0.839505999999999\\
	10	0.864934000000001\\
	11	0.892369999999997\\
	12	0.915543\\
	13	0.969853999999998\\
	14	0.998695999999999\\
	15	1.122178\\
	16	1.144281\\
	17	1.215846\\
	18	1.29345\\
	19	1.411897\\
	20	1.557517\\
};\label{homohetero_solvetime}

\addplot [color=Peach,solid,line width=1.0pt,mark=square,mark options={solid},forget plot]
table[row sep=crcr]{%
	1	0.27826334\\
	2	0.3675772731\\
	3	0.4837504105\\
	4	0.532878877\\
	5	0.6793813546\\
	6	0.7110408441\\
	7	0.8456619088\\
	8	0.9794497247\\
	9	1.0238687355\\
	10	1.0271849938\\
	11	1.0723468167\\
	12	1.1382044402\\
	13	1.1562324274\\
	14	1.2039779611\\
	15	1.2516224987\\
	16	1.2707075564\\
	17	1.3137186648\\
	18	1.3878833799\\
	19	1.4462636452\\
	20	1.4838021216\\
};\label{ADMM1iter_solvetime}

\addplot [color=darktangerine,solid,line width=1.0pt,mark=diamond,mark options={solid},forget plot]
table[row sep=crcr]{%
	1	0.25826334\\
	2	0.39826334\\	
	3	0.64826334\\
	4	0.913527\\
	5	1.538683\\
	6	2.659711\\
	7	4.160772\\
	8	7.833544\\
	9	10.241738\\
	10	15.165793\\
	11	21.415494\\
	12	28.283102\\
	13	38.559402\\
	14	51.979925\\
	15	88.284425\\
	16	144.423943\\
	17	218.711294\\
	18	281.82409\\
	19	348.267213\\
	20	450.46649\\
};\label{heteroedge_decomp_solvetime}

\addplot [color=Red,solid,line width=1.0pt,mark=triangle,mark options={solid},forget plot]
table[row sep=crcr]{%
	1	0.25826334\\
	2	0.39826334\\	
	3	1.02855900000\\
	4	2.05469600000\\
	5	8.50110200000\\
	6	22.19478500000\\
	7	78.74202800000\\
	8	236.19059300000\\
	9	393.71386000000\\
	10	813.78094900000\\
	11	2377.14288100000\\
	12	4783.48893400000\\
	13	20037.98245500001\\
};\label{heteroedge_centralized_blkdiagpersys_solvetime}

\end{axis}
\end{tikzpicture}%
			\end{center}
		\end{subfigure}
	\hspace{0.5cm}
	\begin{subfigure}{0.23\textwidth}
		\ref{homo_Numvar} Homogeneous system \\ 
		\hspace{0.1cm} (Proposition~\ref{prop:smallLMIs_homo}) \\[0.1cm]
	    \ref{homohetero_Numvar} $\aalpha$-$\intercN$-Heterogeneous \\ system 
		(Proposition~\ref{prop:smallLMIs_homohetero}) \\[0.1cm]
		\ref{ADMM1iter_Numvar} Heterogeneous system \\
		(1 iteration of ADMM  \\  
		Algorithm~\ref{alg:ADMM} 
		for 1 subsystem $i$) \\[0.1cm]
		\ref{heteroedgedecomp_Numvar} Heterogeneous system \\
		(Proposition~\ref{prop:smallLMIs_hetero}) \\[0.1cm]
		\ref{heteroedge_centralized_blkdiagpersys_solvetime} Heterogeneous system \\ 
		(Theorem~\ref{the:sproc} with block-diagonal \\ 
		multipliers per subsystem)  \label{fig:scalestudy}
	\end{subfigure}
		\caption{\small 
			Number of optimization variables and solver times versus the number of subsystems $N$, the number of groups $\aalpha$, 
			or the number of neighboring subsystems $\vert \mathcal{N}_i \vert$. 
		\label{fig:scalestudy}	}
	\end{centering}
	\end{scriptsize} 
\end{figure*}

\section{Conclusion and Discussion}

We have presented a scalable distributed controller synthesis for large-scale heterogeneous systems that are interconnected over directed graphs. The interconnection topology of the designed controller may also be directed and different from the topology of the system. 
The scalability of the control design is achieved through a decomposition of the synthesis conditions and through a distributed solution method. Applying the full block S-procedure allowed us to decompose the controller synthesis equations into smaller ones of the size of the individual subsystems. We proposed a distributed synthesis method based on an ADMM scheme with only nearest-neighbor bidirectional communication and without central coordination. 

We further introduced a new system classification as $\aalpha$-$\intercN$-heterogeneous systems, that consist of $\aalpha$ groups of homogeneous subsystems with $\intercN$ different interconnection types. We showed how, based on this classification, the interconnected system model can be transformed to a more compact model, in the case of small values of $\aalpha$ and $\intercN$. 
For these special classes of systems, less conservatism is introduced by using a controller synthesis based on the more compact model that exploits the information of the interconnection topology of the subsystems. 
Numerical examples have illustrated the convergence of the distributed synthesis and the computational scalability of the decomposed synthesis methods.

\section*{Acknowledgement}
The authors would like to thank Dr.\ Goran Banjac from the Automatic Control Laboratory, ETH Zurich for  valuable technical discussions on the topic.

\appendices
\section{System and Controller Matrices for the Interconnected System Model}
\label{sub:LFTMatrices_hetero}

\subsection{System Matrices of \eqref{eq:interconn_openloop_i} for Heterogeneous Systems} \label{subsub:LFTMatrices_hetero} 
The system matrices corresponding to the interconnection channel of subsystem $i$ for a heterogeneous system in \eqref{eq:interconn_openloop_i} are given by 
\begin{equation}
\label{eq:interconn_matrices_hetero}
\begin{aligned}
{B}_{\intercPi, i} &=   \concatIndtwo{k \in \plant{\mathcal{N}}_i}{}{   \small \bma{@{} c@{\,\,\,\,\,\,} c@{}}{ 
		{A}_{ ik}  & {B}_{ \perfi, ik}
}^\trans}^\trans   ,\\
{D}_{\perfo \intercPi, i} &=  \concatIndtwo{k \in \plant{\mathcal{N}}_i}{}{   
	\small  \bma{@{}c@{\,\,\,\,\,\,} c@{}}{ {C}_{\perfo ,  ik}  & {D}_{\perfo \perfi, ik} }^\trans}^\trans , \\
{D}_{y \intercPi, i} &=   \concatIndtwo{k \in \plant{\mathcal{N}}_i}{}{ \small \bma{@{} c@{\,\,\,\,\,\,} c@{}}{ 
		{C}_{y,  ik}  & {D}_{y\perfi, ik}
}^\trans}^\trans,\\
{C}_{\intercPo, i} &=   \concatIndtwo{k \in \plant{\mathcal{N}}_i}{}{ \small \bma{@{} c@{\,\,\,\,\,\,} c@{}}{ 
		I_{n_{x_{k}}}  & 0
}^\trans}   , \\
{D}_{\intercPo \perfi, i} &=   \concatIndtwo{k \in \plant{\mathcal{N}}_i}{}{\small \bma{@{}c@{\,\,\,\,\,\,}c@{}}{ 
		0  & I_{n_{\perfi_{k}}}}^\trans}.
\end{aligned}
\end{equation}

\subsection{System Matrices of \eqref{eq:interconn_openloop_i} for Homogeneous Systems} \label{subsub:LFTMatrices_hetero}

The system matrices corresponding to the interconnection channel of subsystem $i$ in  \eqref{eq:interconn_openloop_i} for a homogeneous system are given by 
\begin{equation}
\label{eq:interconn_matrices_homo}
\begin{aligned}
{B}_{\intercPi, i} &=  { \small \bma{@{}c@{\,\,\,\,\,\,} c@{}}{ 
		{A}_{ ik}   & {B}_{\perfi, ik}
}}    ,\\
{D}_{\perfo \intercPi, i} &=  {  
	\small  \bma{@{}c@{\,\,\,\,\,\,}  c@{}}{ {C}_{\perfo  , ik} &  {D}_{\perfo \perfi, ik} }}   , \\
{D}_{y \intercPi, i} &=   { \small \bma{@{}c@{\,\,\,\,\,\,} c@{}}{ 
		{C}_{y,  ik}  & {D}_{y {w}, ik}
}} ,\\
{C}_{\intercPo, i} &=   { \small \bma{@{}c@{\,\,\,\,\,\,} c@{}}{ 
		I_{n_{x}}  & 0
	}^\trans}   , \\
{D}_{\intercPo \perfi, i} &=   {\small \bma{@{}c@{\,\,\,\,\,\,} c@{}}{ 
		0  & I_{n_{\perfi}}}^\trans}.
\end{aligned}
\end{equation}

\subsection{System Matrices of \eqref{eq:interconn_openloop_i} for $\aalpha$-$\intercN$-Heterogeneous Systems} \label{subsub:LFTMatrices_hetero}

The system matrices corresponding to the interconnection channel of subsystem $i$ in  \eqref{eq:interconn_openloop_i}   
for $\aalpha$ groups of homogeneous subsystems with $\intercN$ different homogeneous interconnections are given by 
\begin{equation}
\label{eq:interconn_matrices_homohetero}
\begin{aligned}
{B}_{\intercPi, i} &=   \sum_{k=1}^{\plant{\intercN}}\nolimits \left(  \kronecker{e_{k}^\trans}{ \small \bma{@{}c@{\,\,\,\,\,\,} c@{}}{ 
		{A}_{ ik}   & {B}_{ \perfi, ik}
}} \right)   ,\\
{D}_{\perfo \intercPi, i} &=  \sum_{k=1}^{\plant{\intercN}}\nolimits \left(  \kronecker{e_{k}^\trans}{   
	\small  \bma{@{}c@{\,\,\,\,\,\,} c@{}}{ {C}_{\perfo ,  ik}  & {D}_{\perfo \perfi, ik} }} \right)  , \\
{D}_{yp ,i} &=   \sum_{k=1}^{\plant{\intercN}}\nolimits \left(  \kronecker{e_{k}^\trans}{ \small \bma{@{}c@{\,\,\,\,\,\,} c@{}}{ 
		{C}_{y,  ik}  & {D}_{y \perfi, ik}
}} \right),\\
{C}_{\intercPo, i} &=   \sum_{k=1}^{\plant{\intercN}}\nolimits  \left( \kronecker{e_{k}}{ \small \bma{@{}c@{\,\,\,\,\,\,} c@{}}{ 
		I_{n_{x_k}}  & 0
	}^\trans} \right)  , \\
{D}_{\intercPo \perfi, i} &=   \sum_{k=1}^{\plant{\intercN}}\nolimits \left(  \kronecker{e_{k}}{\small \bma{@{}c@{\,\,\,\,\,\,}c@{}}{ 
		0  & I_{n_{\perfi_k}}}^\trans} \right).
\end{aligned}
\end{equation}

\subsection{Controller Matrices of \eqref{eq:dofbd} for Interconnected Heterogeneous Controllers} 
\label{subsub:LFTMatrices?controller_hetero}
The controller matrices corresponding to the interconnection channel of subcontroller $i$ for a heterogeneous controller in \eqref{eq:dofbd} are given by 
\begin{equation}
\label{eq:interconn_matrices_controller_hetero}
\begin{aligned}
{B}_{\intercKi, i} &=   \concatIndtwo{k \in \cont{\mathcal{N}}_i}{}{   \small \bma{@{} c@{\,\,\,\,\,\,} c@{}}{ 
		{\AK}_{ ik}  & {\BK}_{ ik}
}^\trans}^\trans   ,\\
{C}_{\intercKi, i} &=  \concatIndtwo{k \in \cont{\mathcal{N}}_i}{}{   
	\small  \bma{@{}c@{\,\,\,\,\,\,} c@{}}{ {\CK}_{ ik}  & {\DK}_{ ik} }^\trans}^\trans , \\
{C}_{\intercKo, i} &=   \concatIndtwo{k \in \cont{\mathcal{N}}_i}{}{ \small \bma{@{} c@{\,\,\,\,\,\,} c@{}}{ 
		I  & 0
}^\trans},\\
{D}_{\intercKo, i} &=   \concatIndtwo{k \in \cont{\mathcal{N}}_i}{}{ \small \bma{@{} c@{\,\,\,\,\,\,} c@{}}{ 
		0  & I
}^\trans}. 
\end{aligned}
\end{equation}
The matrices for the interconnection channel of homogeneous interconnected controllers, or of groups of homogeneous subcontrollers with homogeneous interconnections are analogous to the ones of the system matrices.

\subsection{System Matrices of the Closed Loop in \eqref{eq:closed_loop_PPK_i}}
\label{eq:interconn_matrices_closedloop_hetero}

With the interconnected system and controller realizations in \eqref{eq:interconn_matrices_hetero} and \eqref{eq:interconn_matrices_controller_hetero}, respectively, for a heterogeneous system the closed-loop matrices for subsystem $i$ are the following.
\begin{equation}\label{eq:interconn_matrices_clp_hetero}
\begin{aligned}
{\mathcal{A}}_i &= \bma{@{}c@{\,\,\,\,\,\,}c@{}}{{A}_{i} + {B}_{ u, i} {\DK}_{i} {C}_{y, i} & {B}_{ u, i} {\CK}_i \\ {\BK}_i {C}_{y, i} & {\AK}_i }, \\
{\mathcal{B}}_{1,i} &= \bma{@{}c@{}}{{B}_{\perfi, i} + {B}_{u, i} {\DK}_i {D}_{y \perfi, i} \\ {\BK}_i {D}_{y \perfi, i} }, \\
{\mathcal{C}_{1,i}} &= \bma{@{}c@{\,\,\,\,\,\,}c@{}}{{C}_{\perfo, i} + {D}_{\perfo u, i} {\DK}_i {C}_{y, i} & {D}_{\perfo u, i} {\CK}_i }, \\
{\mathcal{D}_{11,i}} &= \bma{@{}c@{}}{{{D}_{\perfo \perfi, i} + {D}_{\perfo u, i} {\DK}_i {D}_{y \perfi, i}}}, \\
{\mathcal{B}_{2,i}} &= \bma{@{}c@{\,\,\,\,\,\,}c@{}}{{B}_{\intercPi, i} & {B}_{ u, i} C_{\intercKi, i}\\ 0 & B_{\intercKi, i}}, \\
{\mathcal{C}_{2,i}} &= \bma{@{}c@{\,\,\,\,\,\,}c@{}}{{C}_{\intercPo, i} &  0  \\  C_{y, i} {D}_{\intercKo, i} & {C}_{\intercKo, i} }, \\
{\mathcal{D}_{21,i}} &= \bma{@{}c@{}}{{D}_{\intercPo \perfi, i}  \\  {D}_{y \perfi, i} {D}_{\intercKo, i} }, \\
{\mathcal{D}_{12,i}} &= \bma{@{}c@{\,\,\,\,\,\,}c@{}}{{D}_{\perfo \intercPi, i} & {D}_{\perfo u, i} C_{\intercKi, i}}, \\
{\mathcal{D}_{22,i}} &= 0. 
\end{aligned}
\end{equation}

\section{Derivation of Algorithm \ref{alg:ADMM}}
\label{sec:derivationADMM}

\subsection{Distributed Synthesis With Global Coordination}

We start by formulating the classical consensus ADMM problem as in \cite{Boyd2010} and then derive the update steps of Algorithm~\ref{alg:ADMM}. 
With the global variable $\gl$ and the local variables $\y_i$ for subsystems $i$ as defined in Section~\ref{sec:DistributedDesign}, 
and further defining the selection matrices $H_i$ and $E_i$ such that the entries of $H_i \, \gl$ correspond to the local variables of $E_i \y_i$, we formulate the following global consensus constraints 
$$E_i \y_i = H_i \, \gl, ~~ \forall i \in \mathcal{N}. $$

The decomposed synthesis problem with global consensus is then formulated as 
\begin{equation} \label{eq:decomposedsyncentral}
\begin{aligned}
\min_{\y_i} 
& \sum_{i=1}^{N} \big( f_i(\y_i) + g_i(\y_i) 
\big), \\
\mathrm{s.t.} ~~ & E_i \y_i = H_i \, \gl, ~~ \forall i \in \mathcal{N},  
\end{aligned}
\end{equation}
with $f_i(\cdot)$ as defined before and  
$$g_i(\y_i) = 
\mathcal{I}_{\eqref{eq:primalLMIs_small_mult_hetero}}(\y_i) +  \mathcal{I}_{\eqref{eq:primalLMIs_small_hetero}}(\y_i), ~~ \forall i \in \mathcal{N}.$$

The local augmented Lagrangian of subsystem $i$ for the synthesis problem is given as 
\begin{equation}
\begin{aligned}
\mathcal{L}_{\rho,i} = &f_i(\y_i) + g_i(\y_i)
\\ & + \lambda_{{i}}^\trans \left(E_i \y_{i} - H_i \, {\gl} \right) + \frac{\rho}{2}   \| E_i \y_{i} - H_i \, {\gl} \|_2^2. 
\end{aligned}
\end{equation}
The consensus ADMM as in \cite{Boyd2010} involves the following update steps. 
			\begin{equation*}
			\begin{aligned}
			\y_i^{(\kappa+1)}  & = \underset{\y_i}
			{\mathrm{argmin}}  ~ \mathcal{L}_{\rho,i} (\lambda_{i}^{(\kappa)}, \gl^{(\kappa)}, {\y}_i), ~ \forall i \in \mathcal{N}, \\
		{\gl}^{(\kappa+1)} & = \underset{{\gl}}{\mathrm{argmin}} ~ \sum\limits_{\mathcal{N}} \mathcal{L}_{\rho,i} (\y_i^{(\kappa+1)}, \lambda_{i}^{(\kappa)}, \gl), \\
			\lambda_{i}^{(\kappa+1)} & = \lambda_{i}^{(\kappa)} + \rho ~( E_i \y_i^{(\kappa+1)} - H_i \, {\gl}^{(\kappa+1)}), ~\forall i \in \mathcal{N}. \\
			\end{aligned}
			\end{equation*}
This formulation involves a global consensus of all local variables $E_i \y_i$ with the corresponding parts of the global variable $H_i \, \gl$. If a central instance is available and broadcasting is assumed, these global consensus steps can directly be implemented.

\subsection{Decomposed Synthesis Problem With Local Consensus Variables}

\begin{figure}
	\input{figures/interconnections_consensus_1.tex} 
	\includegraphics[width=0.9\columnwidth]{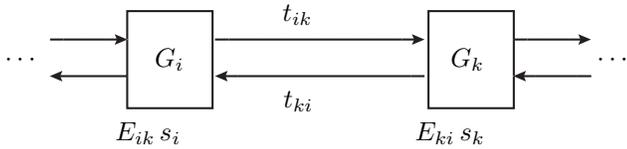} 
	\caption{Local variables per subsystem (here: $i$ and $k$ $\in \mathcal{N}$) and global variables per edge (here: $(i,k)$ $\in \mathcal{E}$). \label{fig:interconnections_consensus}}
\end{figure}

In order to avoid a global coordinator, we aim at eliminating the global consensus variable $\gl$ and therefore introduce the following local consensus variables per interconnection $(i,k)$ and $(k,i)$, 
$t_{ik}$, and $t_{ki}$, 
and corresponding dual multipliers 
$u_{ik}, v_{ik}, u_{ki}$, and $v_{ki}$ 
for subsystems $i$ and $k$, respectively, which allows us to form the consensus constraints 
as shown in Figure~\ref{fig:interconnections_consensus} and Table~\ref{tab:interconnections_consensus}. 

\begin{table}
	\centering 
	\begin{tabular}{cccc}
		\multicolumn{1}{c}{consensus $i$} & \multicolumn{1}{c}{dual var. $i$}  & \multicolumn{1}{c}{consensus $k$} &  \multicolumn{1}{c}{dual var. $k$} \\\hline
		$E_{ik} \, \y_i=t_{ik}$ & $u_{ik}$ & $E_{ki} \, \y_k=t_{ki}$ & $u_{ki}$ \\
		$E_{ki} \, \y_k=t_{ik}$ & $v_{ik}$ & $E_{ik} \, \y_i=t_{ki}$ & $v_{ki}$ 
	\end{tabular}
	\caption{Consensus constraints and corresponding dual variables per subsystem with neighboring subsystems (here $i$ with $k$). }
	\label{tab:interconnections_consensus}
\end{table}

Then, we can formulate the following local synthesis problem for subsystem $i$ with local consensus variables
\begin{equation}\label{eq:localdecomposededges}
\begin{aligned}
&\min_{\y_i, t_{ik}}
 ~~ \sum_{i=1}^{N}  f_i(\y_i) + g_i(\y_i)\\
& ~~~~ \left. 
\begin{aligned}
\mathrm{s.t.} ~ & ~~~~ E_{ik}\, \y_i = t_{ik}, \\
& ~~~~E_{ki} \, \y_k = t_{ik}. 
\end{aligned}
\right\rbrace 
\forall i \in \mathcal{N}, k \in \mathcal{N}_i.
\end{aligned}
\end{equation}

\subsection{Derivation of the ADMM Iterations in Algorithm~\ref{alg:ADMM}} 

Because of the symmetry of the undirected communication graph, \eqref{eq:localdecomposededges} can be transformed into the simplified two-step ADMM algorithm in Algorithm~\ref{alg:ADMM} which will be derived in the following.

For the optimization problem in \eqref{eq:localdecomposededges}, we formulate the augmented Lagrangian as 
\begin{equation}
\begin{aligned}
\mathcal{L}_{\rho}(\y,t,(u,v)) &= \sum_{i=1}^{N} \big( 
f_i(\y_i) + g_i(\y_i)   \\
&+ \sum_{k\in\mathcal{N}_i} \big( 
 u_{ik}^\top ( E_{ik} \, \y_i - t_{ik} ) 
+ \frac{\rho}{2}   \| E_{ik} \, \y_i - t_{ik} \|_2^2  \\
& ~~~~~~ +   v_{ik}^\top ( E_{ki} \, \y_k - t_{ik} )  + \frac{\rho}{2}   \| E_{ki} \, \y_k - t_{ik} \|_2^2   
\big)  
\big),  
\end{aligned}
\end{equation}
where $\y$, $t$, $u$, and $v$ are defined as the stacked vectors $\y = \concatIndtwo{i \in \mathcal{N}}{}{\y_{i}}$, $t = \concatIndtwo{(i,k) \in \mathcal{E}}{}{t_{ik}}$, $u = \concatIndtwo{(i,k) \in \mathcal{E}}{}{u_{ik}}$, and $v = \concatIndtwo{(i,k) \in \mathcal{E}}{}{v_{ik}}$.

The standard consensus ADMM iterations as in \cite{Boyd2010} with respect to $\mathcal{L}_{\rho}(\y,t,(u,v))$ are the following. 
\begin{equation}
\begin{aligned}
\y_i^{(\kappa+1)} \gets & ~\underset{\y_i}{\mathrm{argmin}} \Big \{ f_i(\y_i) + g_i(\y_i) \\ 
& + \sum_{k \in \mathcal{N}_i} 
\Big(  
(u_{ik}^{(\kappa)} + v_{ki}^{(\kappa)})^\top (E_{ik} \, \y_i)
\\
& + \frac{\rho}{2} \|E_{ik} \, \y_i-t_{ik}^{(\kappa)}\|_2^2 
 + \frac{\rho}{2} \|E_{ik} \, \y_i-{t}_{ki}^{(\kappa)}\|_2^2 \Big)  \Big \}, \\
t_{ik}^{(\kappa+1)} \gets & ~\underset{t_{ik}}{\mathrm{argmin}} \Big \{  
- 
t_{ik}^\top (u_{ik}^{(\kappa)} + v_{ik}^{(\kappa)}) 
\\
&+ \frac{\rho}{2} \|t_{ik}-E_{ik} \, \y_i^{(\kappa+1)}\|_2^2 
+ \frac{\rho}{2} \|t_{ik}-E_{ki} \, \y_k^{(\kappa+1)}\|_2^2   \Big \}, \\
u_{ik}^{(\kappa+1)} &\gets %
u_{ik}^{(\kappa)} + \rho \left( E_{ik} \, \y_i^{(\kappa +1)} - t_{ik}^{(\kappa+1)} \right),  \\
v_{ik}^{(\kappa+1)} &\gets %
v_{ik}^{(\kappa)} + \rho \left( E_{ki} \, \y_k^{(\kappa +1)} - t_{ik}^{(\kappa+1)} \right).  \\
\end{aligned}
\end{equation}
The minimization step of $t_{ik}^{(\kappa+1)}$ admits the closed-form solution 
\begin{equation} \label{eq:clsdformsoltik}
t_{ik}^{(\kappa+1)} = \frac{1}{2} \left( E_{ik} \, \y_i^{(\kappa+1)}+ E_{ki} \, \y_k^{(\kappa+1)} \right) + \frac{1}{2 \rho} \left( u_{ik}^{(\kappa)}+v_{ik}^{(\kappa)} \right). 
\end{equation} 
Summing the update equations of $u_{ik}^{(\kappa+1)}$ and $v_{ik}^{(\kappa+1)}$, and replacing $t_{ik}^{(\kappa+1)}$ with the explicit solution in \eqref{eq:clsdformsoltik} leads to 
\begin{equation}\label{eq:8}
u_{ik}^{(\kappa)}+v_{ik}^{(\kappa)} = 0, 
\end{equation} 
and thus, the minimization step of $t_{ik}^{(\kappa+1)}$ simplifies to the following update 
\begin{equation}\label{eq:tik}
t_{ik}^{(\kappa+1)} = \frac{1}{2} \left( E_{ik} \, \y_i^{(\kappa+1)}+ E_{ki} \, \y_k^{(\kappa+1)} \right). 
\end{equation} 
Furthermore, the update step of $u_{ik}^{(\kappa+1)}$ becomes 
\begin{equation}\label{eq:10}
u_{ik}^{(\kappa+1)} \gets u_{ik}^{(\kappa)} + \frac{\rho}{2} \left( E_{ik} \, \y_i^{(\kappa+1)}+ E_{ki} \, \y_k^{(\kappa+1)} \right).
\end{equation}
Note that for initial conditions $t_{ik}^{(0)} = t_{ki}^{(0)}$, it follows from \eqref{eq:tik} that $t_{ik}^{(\kappa)} = t_{ki}^{(\kappa)}$ for all $\kappa > 0$.  
Also, it follows from \eqref{eq:8} and \eqref{eq:10} that for initial conditions $u_{ik}^{(0)} = v_{ik}^{(0)} = 0$ and $u_{ik}^{(0)} = u_{ki}^{(0)} = 0$, then $u_{ik}^{(\kappa)} = -v_{ik}^{(\kappa)}$ and $u_{ik}^{(\kappa)} = -u_{ki}^{(\kappa)}$, for all $\kappa > 0$. 
If we define 
$$
\lambda_i = u_{ik} - v_{ik} = 2 u_{ik}, 
$$ 
this leads to the update 
$$\lambda_{i}^{(\kappa+1)} = \lambda_{i}^{(\kappa)} + \rho \sum_{k \in \mathcal{N}_i} \left( T_{ik} \,  \y_i^{(\kappa)}- T_{ki} \, \y_k^{(\kappa)} \right),$$ and we arrive at the ADMM iterations in Algorithm~\ref{alg:ADMM}, with $T_{ik}$ as defined in Section~\ref{sec:DistributedDesign}.

\begin{myrem}
	In the case of the consensus over $\gammainv_i$ over all subsystems $i$, and the pairwise consensus over the multipliers defined per undirected edge of two neighboring subsystems, the dual multipliers are defined by 
\begin{equation}
\begin{aligned}
\lambda_i &= \left[ \sum_{k \in \mathcal{N}_i}(u_{\gammainv_{ik}}-v_{\gammainv_{ik}}), \concatIndtwo{k \in \mathcal{N}_i}{}{ \left[u_{m_{ik}}^\trans-v_{m_{ik}}^\trans,~ u_{m_{ki}}^\trans-v_{m_{ki}}^\trans \right]} \right]^\trans \\
&= 2 \left[ \sum_{k \in \mathcal{N}_i}u_{\gammainv_{ik}}, \concatIndtwo{k \in \mathcal{N}_i}{}{ \left[u_{m_{ik}}^\trans,~ u_{m_{ki}}^\trans \right]} \right]^\trans,
\end{aligned}
\end{equation}
where the first sum takes care of the consensus over $\gamma_i$ and the remaining parts take care of the consensus over the multipliers of the edges. 
\end{myrem}

\section{Derivation of the residuals in \eqref{eq:residuals}}
\label{subsec:derivationresiduals}
Starting from the definition of the primal and dual residuals \cite{Boyd2010}, we have 
\begin{equation} 
\begin{aligned}
r^{(\iter + 1)} &=  \concatIndtwo{i = 1}{N}{ \concatIndtwo{k \in \mathcal{N}_i}{}{ r_{ik}^{(\iter+1)} } }, \\
d^{(\iter + 1)} &=  \concatIndtwo{i = 1}{N}{ \concatIndtwo{k \in \mathcal{N}_i}{}{ d_{ik}^{(\iter+1)} } }, 
\end{aligned}
\end{equation}
with 
\begin{equation}
\begin{aligned}
r_{ik}^{(\iter+1)} &= E_{ik} \y_i^{(\iter + 1)} - t_{ik}^{(\iter + 1)}, \\
d_{ik}^{(\iter+1)} &= t_{ik}^{(\iter + 1)} - t_{ik}^{(\iter)}. 
\end{aligned}
\end{equation}
Replacing $t_{ik}^{(\iter + 1)}$ and $t_{ik}^{(\iter)}$ by the expressions in \eqref{eq:tik}, we obtain 
\begin{equation} \label{eq:derivation_residuals}
\begin{aligned}
r_{ik}^{(\iter+1)} &= E_{ik} \y_i^{(\iter + 1)} - \frac{1}{2} \left( E_{ik} \y_i^{(\iter + 1)} + E_{ki} \y_k^{(\iter + 1)} \right) , \\
d_{ik}^{(\iter+1)} &= \frac{1}{2} \left( E_{ik} \y_i^{(\iter + 1)} + E_{ki} \y_k^{(\iter + 1)} \right) - \frac{1}{2} \left( E_{ik} \y_i^{(\iter)} + E_{ki} \y_k^{(\iter)} \right),
\end{aligned}
\end{equation}
which leads to \eqref{eq:residuals} in Section~\ref{sec:NumericalExample}.

\if01 
\begin{IEEEbiography}[{\includegraphics[width=1in,height=1.25in,clip,keepaspectratio]{Stuerz_.eps}}]{Yvonne R.\ St\"urz}
	is a Postdoctoral Researcher and Marie-Sk{\l}odowska-Curie Fellow at the Model Predictive Control Laboratory at the University of California, Berkeley, USA, and affiliated with the Automatic Control Department at KTH, Sweden. 
	She received her PhD from the Automatic Control Laboratory at the Swiss Federal Institute of Technology (ETH), Zurich, Switzerland, in 2019. 
	Prior to that, she received the Bachelor's and Master's degrees in mechanical engineering from the Technical University of Munich, Germany in 2012 and 2014, respectively, and the Diploma degree in general engineering science from the Ecole Centrale Paris, France in 2013. 
	From 2008 to 2014, she was awarded scholarships from the German National Academic Foundation and the Ulderup Foundation. She gained industrial experience through internships with Bosch, Porsche and Siemens in Germany and with Siemens SFAE in China.
	Her research interests include distributed and optimal control, learning-based model predictive control with applications to multi-robotic systems and UAVs, large-scale heterogeneous interconnected systems, digital fabrication and autonomous driving. 
\end{IEEEbiography}
\begin{IEEEbiography}[{\includegraphics[width=1in,height=1.25in,clip,keepaspectratio]{Eichler.eps}}]{Annika Eichler}
received the B.Sc. degree in General Engineering Science, in 2008, and the Dipl.-Ing. degree in Mechatronics, in 2011, from the Hamburg University of Technology, Germany, from where she received her Ph.D. degree, in 2015, at the Institute of Control Systems. From 2015 to 2017 she was a Post-Doctoral Researcher at the Automatic Control Laboratory, ETH Zurich, Zürich, Switzerland, where she was working as Senior Researcher until 2019. Currently she is Senior Reseacher at DESY, Hamburg, Germany. Her research interests include distributed, data-driven and robust control with the focus to energy systems, where tools from model predictive control, dynamic programming and robust optimization are applied.
\end{IEEEbiography}
\begin{IEEEbiography}[{\includegraphics[width=1in,height=1.25in,clip,keepaspectratio]{Smith_.eps}}]{Roy S.\ Smith}
is a professor of Electrical Engineering at the Swiss Federal Institute of Technology (ETH), Zürich. Prior to joining ETH in 2011, he was on the faculty of the University of California, Santa Barbara, from 1990 to 2010. His Ph.D. is from the California Institute of Technology (1990) and his undergraduate degree is from the University of Canterbury (1980) in his native New Zealand. He has been a long-time consultant to the NASA Jet Propulsion Laboratory and has industrial experience in automotive control and power system design. His research interests involve the modeling, identification, and control of uncertain systems. Particular control application domains of interest include chemical processes, flexible structure vibration, spacecraft and vehicle formations, aerodynamic control of kites, automotive engines, Mars aeromaneuvering entry design, building energy control, and thermoacoustic machines. He is a Fellow of the IEEE and the IFAC, an Associate Fellow of the AIAA, and a member of SIAM.
\end{IEEEbiography}
\fi






\bibliographystyle{IEEEtran}
\bibliography{IEEEabrv,mybibliography4}

\begin{thebibliography}{10}
\providecommand{\url}[1]{#1}
\csname url@rmstyle\endcsname
\providecommand{\newblock}{\relax}
\providecommand{\bibinfo}[2]{#2}
\providecommand\BIBentrySTDinterwordspacing{\spaceskip=0pt\relax}
\providecommand\BIBentryALTinterwordstretchfactor{4}
\providecommand\BIBentryALTinterwordspacing{\spaceskip=\fontdimen2\font plus
\BIBentryALTinterwordstretchfactor\fontdimen3\font minus
  \fontdimen4\font\relax}
\providecommand\BIBforeignlanguage[2]{{%
\expandafter\ifx\csname l@#1\endcsname\relax
\typeout{** WARNING: IEEEtran.bst: No hyphenation pattern has been}%
\typeout{** loaded for the language `#1'. Using the pattern for}%
\typeout{** the default language instead.}%
\else
\language=\csname l@#1\endcsname
\fi
#2}}

\bibitem{DAndrea2003}
R.~D'Andrea and G.~E. Dullerud, ``{Distributed control design for spatially
  interconnected systems},'' \emph{IEEE Trans. Automat. Contr.}, vol.~48,
  no.~9, pp. 1478--1495, 2003.

\bibitem{Langbort2004}
C.~Langbort, R.~S. Chandra, and R.~D'Andrea, ``{Distributed control design for
  systems interconnected over an arbitrary graph},'' \emph{IEEE Trans. Automat.
  Contr.}, vol.~49, no.~9, pp. 1502--1519, 2004.

\bibitem{Murray2007}
R.~M. Murray, ``{Recent research in cooperative control of multivehicle
  systems},'' \emph{J. Dyn. Syst. Meas. Control}, vol. 129, no.~5, p. 571,
  2007.

\bibitem{Smith2007}
R.~S. Smith and F.~Y. Hadaegh, ``{Closed-loop dynamics of cooperative vehicle
  formations with parallel estimators and communication},'' \emph{IEEE Trans.
  Automat. Contr.}, vol.~52, no.~8, pp. 1404--1414, 2007.

\bibitem{Stuerz2017a}
Y.~R. St\"urz, A.~Eichler, and R.~S. Smith, ``{A framework for distributed
  control based on overlapping estimation for cooperative tasks},'' in
  \emph{IFAC World Congr.}, vol.~50, no.~1.\hskip 1em plus 0.5em minus
  0.4em\relax Elsevier B.V., 2017, pp. 14\,296--14\,301.

\bibitem{Alam2011}
A.~Alam, A.~Gattami, and K.~Johansson, ``{Optimal distributed controller
  synthesis for chain structures applications to vehicle formations},'' in
  \emph{IEEE Conf. Decis. Control}, 2011, pp. 6894--6900.

\bibitem{Gahinet1995}
P.~Gahinet, ``{A convex characterization of gain-scheduled $\mathcal{H}_\infty$
  controllers},'' \emph{IEEE Trans. Automat. Contr.}, vol.~40, no.~5, pp.
  853--864, 1995.

\bibitem{Scorletti1997}
G.~Scorletti and G.~Duc, ``{A convex approach to decentralized
  $\mathcal{H}_\infty$ control},'' in \emph{Am. Control Conf.}, 1997, pp.
  2390--2394.

\bibitem{Dullerud2004}
G.~E. Dullerud and R.~D'Andrea, ``{Distributed control of heterogeneous
  systems},'' \emph{IEEE Trans. Automat. Contr.}, vol.~49, no.~12, pp.
  2113--2128, 2004.

\bibitem{Langbort2003}
C.~Langbort and R.~D'Andrea, ``{Distributed control of heterogeneous
  systems},'' in \emph{IEEE Conf. Decis. Control}, 2003, pp. 2835--2840.

\bibitem{Rice2009}
J.~K. Rice and M.~Verhaegen, ``{Distributed control: A sequentially
  semi-separable approach for spatially heterogeneous linear systems},''
  \emph{IEEE Trans. Automat. Contr.}, vol.~54, no.~6, pp. 1270--1283, 2009.

\bibitem{Borrelli2008}
F.~Borrelli and T.~Keviczky, ``{Distributed LQR design for identical
  dynamically decoupled systems},'' \emph{IEEE Trans. Automat. Contr.},
  vol.~53, no.~8, pp. 1901--1912, 2008.

\bibitem{Rantzer1997}
A.~Rantzer and A.~Megretski, ``{System analysis via integral quadratic
  constraints},'' \emph{IEEE Trans. Automat. Contr.}, vol.~42, no.~6, pp.
  819--830, 1997.

\bibitem{Meinsma2000}
G.~Meinsma, T.~Iwasaki, and M.~Fu, ``{When is (D,G)-scaling both necessary and
  sufficient},'' \emph{IEEE Trans. Automat. Contr.}, vol.~45, no.~9, pp.
  1755--1759, 2000.

\bibitem{Scherer2001}
\BIBentryALTinterwordspacing
C.~Scherer, ``{Theory of robust control},'' \emph{Delft Univ. Technol.}, pp.
  1--160, 2001. [Online]. Available:
  \url{http://www.imng.uni-stuttgart.de/mst/robust/RCNotes.pdf}
\BIBentrySTDinterwordspacing

\bibitem{Massioni2009}
P.~Massioni and M.~Verhaegen, ``{Distributed control for identical dynamically
  coupled systems: A decomposition approach},'' \emph{IEEE Trans. Automat.
  Contr.}, vol.~54, no.~1, pp. 124--135, 2009.

\bibitem{Massioni2014}
P.~Massioni, ``{Distributed control for alpha-heterogeneous dynamically coupled
  systems},'' \emph{Syst. Control Lett.}, vol.~72, pp. 30--35, 2014.

\bibitem{Massioni2010}
------, ``{Decomposition methods for distributed control and identification},''
  Ph.D. dissertation, Technical University Delft, 2010.

\bibitem{Hoffmann2015}
C.~Hoffmann, A.~Eichler, and H.~Werner, ``{Control of heterogeneous groups of
  systems interconnected through directed and switching topologies},''
  \emph{IEEE Trans. Automat. Contr.}, vol.~60, no.~7, pp. 1904--1909, 2015.

\bibitem{Hoffmann2016}
C.~Hoffmann and H.~Werner, ``{Convex distributed controller synthesis for
  interconnected heterogeneous subsystems via virtual normal interconnection
  matrices},'' \emph{IEEE Trans. Automat. Contr.}, vol.~62, no.~10, pp. 5337 --
  5342, 2017.

\bibitem{Stuerz2018a}
Y.~R. St\"urz, A.~Eichler, and R.~S. Smith, ``{Scalable controller synthesis
  for heterogeneous interconnected systems applicable to an overlapping control
  framework},'' in \emph{Eur. Control Conf.}, vol.~1, 2018, pp. 2561--2568.

\bibitem{Langbort2004a}
C.~Langbort, R.~D'Andrea, and S.~Boyd, ``{A decomposition approach to
  distributed analysis of networked systems},'' in \emph{IEEE Conf. Decis.
  Control}, 2004, pp. 3980--3985.

\bibitem{Viccione2009}
P.~Viccione, C.~W. Scherer, and M.~Innocenti, ``{LPV synthesis with integral
  quadratic constraints for distributed control of interconnected systems},''
  in \emph{6th IFAC Symp. Robust Control Des.}, vol.~42, no.~6.\hskip 1em plus
  0.5em minus 0.4em\relax IFAC, 2009, pp. 13--18.

\bibitem{Zheng2018a}
Y.~Zheng, R.~P. Mason, and A.~Papachristodoulou, ``{Scalable design of
  structured controllers using chordal decomposition},'' \emph{IEEE Trans.
  Automat. Contr.}, vol.~63, no.~3, pp. 752--767, 2018.

\bibitem{Boyd2010}
S.~Boyd, N.~Parikh, B.~P. {E Chu}, and J.~Eckstein, ``{Distributed optimization
  and statistical learning via the alternating direction method of
  multipliers},'' \emph{Found. Trends Mach. Learn.}, vol.~3, no.~1, pp. 1--122,
  2010.

\bibitem{Molzahn2017}
D.~K. Molzahn, F.~D{\"{o}}rfler, H.~Sandberg, S.~H. Low, S.~Chakrabarti,
  R.~Baldick, and J.~Lavaei, ``{A survey of distributed optimization and
  control algorithms for electric power systems},'' \emph{IEEE Trans. Smart
  Grid}, vol.~8, no.~6, pp. 2941--2962, 2017.

\bibitem{Braun2018}
P.~Braun, T.~Faulwasser, L.~Gr{\"{u}}ne, C.~M. Kellett, S.~R. Weller, and
  K.~Worthmann, ``{Hierarchical distributed ADMM for predictive control with
  applications in power networks},'' \emph{IFAC J. Syst. Control}, vol.~3, pp.
  10--22, 2018.

\bibitem{Guo2018}
J.~Guo, G.~Hug, and O.~Tonguz, ``{Impact of communication delay on asynchronous
  distributed optimal power flow using ADMM},'' \emph{Int. Conf. Smart Grid},
  pp. 177--182, 2017.

\bibitem{Hansson2014}
A.~Hansson and M.~Verhaegen, ``{Distributed system identification with ADMM},''
  \emph{IEEE Conf. Decis. Control}, pp. 290--295, 2015.

\bibitem{Stuerz2016b}
Y.~R. St\"urz, M.~Morari, and R.~S. Smith, ``{Two methods for the
  identification of uncertain parameters of an architectural cable net
  geometry},'' in \emph{IEEE Conf. Control Appl.}, 2016, pp. 804--809.

\bibitem{Ba2015}
Q.~Ba, K.~Savla, and G.~Como, ``{Distributed optimal equilibrium selection for
  traffic flow over networks},'' \emph{IEEE Conf. Decis. Control}, pp.
  6942--6947, 2015.

\bibitem{Dhingra2014}
N.~K. Dhingra, M.~R. Jovanovic, and Z.~Q. Luo, ``{An ADMM algorithm for optimal
  sensor and actuator selection},'' \emph{IEEE Conf. Decis. Control}, pp.
  4039--4044, 2015.

\bibitem{Graf2019}
P.~Graf, J.~Annoni, C.~Bay, D.~Biagioni, D.~Sigler, M.~Lunacek, and W.~Jones,
  ``{Distributed reinforcement learning with ADMM-RL},'' \emph{Proc. Am.
  Control Conf.}, pp. 4159--4166, 2019.

\bibitem{Alonso2019}
C.~A. Alonso and N.~Matni, ``{Distributed and localized model predictive
  control via system level synthesis},'' 2019.

\bibitem{Mateos2010}
G.~Mateos, J.~A. Bazerque, and G.~B. Giannakis, ``{Distributed sparse linear
  regression},'' \emph{IEEE Trans. Signal Process.}, vol.~58, no.~10, pp.
  5262--5276, 2010.

\bibitem{Chang2015}
T.~H. Chang, M.~Hong, and X.~Wang, ``{Multi-agent distributed optimization via
  inexact consensus ADMM},'' \emph{IEEE Trans. Signal Process.}, vol.~63,
  no.~2, pp. 482--497, 2015.

\bibitem{Banjac2018}
G.~Banjac, F.~Rey, P.~Goulart, and J.~Lygeros, ``{Decentralized resource
  allocation via dual consensus ADMM},'' \emph{Amer. Contr. Conf.}, pp.
  2789--2794, 2019.

\bibitem{Darup2019}
M.~S. Darup, G.~Book, and P.~Giselsson, ``{Towards real-time ADMM for linear
  MPC},'' \emph{Eur. Control Conf.}, pp. 4276--4282, 2019.

\bibitem{Stuerz2017b}
Y.~R. St\"urz, A.~Eichler, and R.~S. Smith, ``{Fixed mode elimination by
  minimum communication within an estimator-based framework for distributed
  control},'' \emph{IEEE Control Syst. Lett.}, vol.~1, no.~2, pp. 346--351,
  2017.

\end{thebibliography}

\end{document}